\documentclass[12pt]{article}%
\usepackage{geometry}
\usepackage{amsmath}
\usepackage{amsfonts}
\usepackage{amssymb}
\usepackage{graphicx}
\usepackage{bm}
\usepackage{authblk}
\usepackage[htt]{hyphenat}%
\providecommand{\U}[1]{\protect\rule{.1in}{.1in}}
\begin{document}
\title{
\vspace{-2cm}\hfill{\normalsize NSF-KITP-11-092}\\[1cm]
 Dynamical aspects of inextensible chains}
\author[1,2]{Franco Ferrari \thanks{ferrari@fermi.fiz.univ.szczecin.pl}}
\affil[1]{Institute of Physics and CASA*,
    University of Szczecin,\\  ul. Wielkopolska 15, 70-451 Szczecin,
    Poland}
\affil[2]{Kavli Institute for Theoretical Physics, University of
  California, Santa Barbara, California 93106, USA}
\author[3]{Maciej Pyrka \thanks{maciej-p@gumed.edu.pl}}

\affil[3]{Department of Physics and Biophysics, Medical University of
  Gda\'nsk, ul. D\c{e}binki 1, 80-211 Gda\'nsk}
%\author{and Maciej Pyrka\thanks{maciej-p@gumed.edu.pl}}
%{\it Department of Physics and Biophysics}
%}
\date{\today}
\maketitle

%\author{Franco Ferrari and Maciej Pyrka\\\textit{Institute of Physics
%    and CASA*, University of Szczecin,}\\\textit{ul. Wielkopolska 15,
%    70-451 Szczecin, Poland }} 

\begin{abstract}

In the present work the dynamics of a continuous inextensible
chain is studied.
The chain is regarded as a system of small particles subjected to
constraints on their reciprocal distances.
It is proposed a treatment of systems of this kind based
on a set
Langevin equations in which the noise is characterized by
a non-gaussian probability distribution.
The method is explained in the case of a freely hinged chain.
In particular, the generating functional of the correlation functions of
the relevant degrees of freedom which describe the conformations of
this chain is derived.
It is shown that in the continuous limit this generating functional
coincides with a model of an inextensible chain previously discussed
by one of the authors of this work.
Next, the approach developed here is applied to a inextensible chain,
called the freely jointed bar chain,
in which the basic units are small extended objects.
The generating functional 
of the freely jointed bar chain is constructed. It is shown that it
differs profoundly from that of the freely hinged chain. Despite the
differences, it is verified that in the
continuous limit both
generating functionals coincide as it is expected.
\end{abstract}

\section{Introduction}
There are several physical situations
in which it is required that a polymer chain is inextensible.
This happens for instance
 when the 
chain is 
pulled at its ends by strong forces \cite{febbo,marko1,nelson,
  marko2,liverpool} in experiments of DNA micromanipulations.
In order to impose the constraints in
the stochastic equations governing the motion of the chain
it is possible to apply
a variety of powerful techniques \cite{fixman, blundell,Gomes,Mochizuki,
  okano, Muthukumar, Hinch, curtiss}.
Progresses in the understanding of statistic and
dynamical aspects  of inextensible chains have been obtained mainly
with the help of 
numerical simulations, see for example
Refs.~\cite{Doyle,article2, abraham, montesi, klaveness,
  butler, echenique, peters}. 
For analytical calculations,  models in which the condition of
inextensibility is relaxed are usually preferred.
This is the case of the
gaussian chain \cite{doiedwards}, which 
often replaces the freely jointed chain \cite{kramers} in the
study of the statistical behavior of polymers systems.
Other examples of
theories in which the length of the chain is allowed to become
infinitely long are the models of Rouse \cite{rouse} and Zimm 
\cite{zimm}. They provide a 
satisfactory description of the dynamics of polymer chains in
solutions, but could fail in the case of a chain that is stretched
under strong external forces applied at its ends.

For these reasons it is highly desirable to construct a path integral
model describing the dynamics of an  inextensible chain
which is simple enough to allow analytic calculations.
In particular, we are interested in the derivation of the generating
functional of the correlation functions of the relevant variables that
describe the conformation of the chain.
The usual starting point is a 
 discrete mechanical model which
provides a coarse grained approximation of the chain.
Successively,
a continuous
limit is performed, in which the number of elements becomes infinitely
large and 
their length approaches zero, while the total length of the chain $L$
is kept constant. This limit is useful
in order to smooth out 
the dependence on the details of the particular mechanical model
chosen.
In the first part of this article the case of a freely hinged chain
(FHC) is considered. 
This is a system of beads connected together by massless links of
fixed length $a$.
The  FHC
has been studied in the past  in
Refs.~\cite{EdwGoo,EdwGoo2,EdwGoo3} in connection  with the dynamics
of cold polymers or a gas of hot polymers.
In Ref.~\cite{FePaVi} it has been shown that the generating functional
for a FHC coincides in the continuous limit
with the partition function
of a theory which closely resembles a nonlinear sigma model. For this
reason it has been called the generalized 
nonlinear sigma 
model or simply GNL$\sigma$M. Within the
GNL$\sigma$M it is possible to perform analytical calculations
of physical quantities that can be applied  to study
the dynamics of cold chains or of chains moving in a very viscous
environment \cite{FePaVi,FePaVi2,FePaVi3}. 
One drawback of the GNL$\sigma$M is that it regards
the FHC as a gas
of fluctuating particles with 
the inextensibility constraints being implemented
by means of a product of delta functions.
The introduction of delta functions to fix constraints is a standard
procedure in the investigations of the statistical mechanics of
polymers, where it is an useful tool in order to limit the
conformations of the chain. 
In the case 
of dynamics, however, a physical explanation
 of the appearance of the delta 
functions inside the path integral of the GNL$\sigma$M is necessary,
otherwise the 
connection with the
stochastic process of the fluctuating chain is lost.
Such connection was up to now somewhat obscure, despite the fact that
in Ref.~\cite{FePaVi4} it has been possible to verify that
the GNL$\sigma$M
reproduces as expected the sum over all chain conformations
satisfying a free
Langevin equation and
subjected to constraints that ensure the property of inextensibility.

One goal of this work is to provide a  physical interpretation
of the GNL$\sigma$M as a
stochastic process in which the  noise is non-gaussian.
The starting point is the observation that the chain  degrees of
freedom  which are constrained by the inextensibility requirement
are frozen and thus cannot be influenced
by the noise. This implies that the noise must be constrained too.
The conditions
on the noise are imposed in the present approach 
in a soft way using an elastic potential. It is the
introduction of this 
potential that modifies the noise distribution and makes it non-gaussian.
The rigid constraints are recovered in
 the limit in which the elastic constants become
infinite. 
Following this strategy,
the motion of the beads of the chain is described by
 a system of free Langevin equations in
which the random forces
are characterized by a non-gaussian distribution. For any finite value of
the elastic constants, the related  generating
functional corresponds to that of a chain consisting of beads
connected together by springs.
With increasing values of the elastic constants, the springs become stiffer
and stiffer until at the end an inextensible chain is obtained.
Let us note that also the beads of the Rouse model
are joined together by springs. However, in that model the potentials
are chosen in such a way that the springs are in their rest positions
only when the distances between neighboring beads is zero. For this
reason, in the limit of infinite elastic constants, the chain collapses
to a point. Other differences of the GNL$\sigma$M from the
Rouse model have been discussed in  details in Ref.~\cite{FePaVi}.
The present strategy to enforce the inextensibility constraints
with the help of potentials  
closely resembles the way in  which 
stiff constraints (sometimes called flexible constraints)
are imposed \cite{echenique,peters} in stochastic equations. 
Stiff constraints are 
fixed in fact with the help of potentials
\cite{warner,fan} whose
role is to create an energy barrier 
that strongly suppresses those  chain conformations that do
not satisfy the constraints. 
The form of these potentials, like for instance the FENE potential, is
usually too complicated to allow 
analytic calculations. In the case of the
GNL$\sigma$M, instead, there are only elastic potentials, which are
relatively simple. Another striking difference with respect to stiff
constraints is that in our approach it is taken the limit in which
the strength of the elastic forces become infinite. In this limit the
constraints become rigid and not flexible.

In the second part of this work the idea of Ref.~\cite{liverpool} is
investigated. According to that idea, a chain may be
realized by connecting together basic inextensible units,
like for instance small bars. In this way, the length of all elements
of the resulting chain is always constant by construction without the
need of imposing the cumbersome 
holonomic constraints that are otherwise necessary to enforce the
property of inextensibility. Constraints are still necessary in order
to connect together the basic units, but they are very simple and
their imposition is straightforward.
In deriving the generating functional for a system of this kind, however, the
fact that the basic inextensible units are not point-like objects, but
rigid bodies with rotational degrees of freedom, is the origin of
several complications.
For this reason, in this work the basic inextensible units are built
starting from a number $K$ of beads of diameter $\Delta l$, called
hereafter spheres. The  motion of the spheres is  
constrained in such a way that they stay aligned along a segment
of fixed length.
The shape of the whole
system formed by the $K$ spheres is
that of a shish-kebab and offers a good
approximation of a bar. 
Let us note that the constraints needed in
order to form the shish-kebabs consist in conditions on the distances
between pairs of points. Thus we can apply the same strategy used to
implement the inextensibility constraints in the FHC, which are of the
same type.
In the limit in which the number of spheres approaches infinity and
their radius becomes vanishingly small, while the length of the basic
units remains constant, one obtains 
what has been called here the freely jointed bar chain or simply FJBC.
This is a discrete  chain composed by
one-dimensional bars with uniform distribution of mass.

The expression of the generating functional of the correlation
functions of the shish-kebab chain has been provided
in the most general case of a chain containing $N$ basic units, each
composed by a number $K$ of beads. 
The generating functional of the FJBC in the limit $K\to+infty$ and
$\Delta l\to 0$
has been derived.
As discussed in
\cite{liverpool}, the constraints that are needed to held together the
basic units are almost trivial and the associated spurious degrees of
freedom can be easily eliminated.
Finally, the continuous limit of the generating functional of the FJBC
has been performed. In this case several simplifications occur and, at
the end, it has been possible to prove that the generating functional
of the resulting continuous system coincides with the
GNL$\sigma$M. This is an expected result, because in the continuous
limit the mechanical details of the underlying discrete chain should
not play any role.

The material presented in this paper is organized as follows.
In Section \ref{section:classical} the classical aspects of the FHC
and the FJBC are analyzed. 
For the interested reader, a more extensive discussion on
this subject can be found in 
Ref.~\cite{pieranski}.
First, the case of the FHC
is briefly 
reviewed.
Next, the FJBC is studied. 
It is shown  that, while the kinetic and
potential  energies of the FJBC differ considerably from those of the
FHC, they 
coincide in the  continuous limit.
The discussion of the classical aspects of the chain dynamics is very
important to establish a mathematically convenient description of the
FJBC, that 
will be used later to study its statistical dynamics. The classical
dynamics of inextensible chains is also very interesting for concrete
applications ranging from cosmetics to computer graphics.
Section \ref{section:statistical} is dedicated to the statistical
dynamics of the chain. In the first part of that Section, 
the
generating functional of the correlation functions of the radius
vectors which specify the positions of the beads of the FHC is derived
in path integral form. The  
constraints  required by the fact that the length of the links
connecting the 
beads is constant are imposed by using elastic potentials as explained
before. When the strength of these potentials becomes infinite, 
 the
 path integral formulation of the generating functional
 of the FHC given in Ref.~\cite{FePaVi} is recovered. 
In the continuous
limit the GNL$\sigma$M is obtained. We fill also a gap 
of our previous publications concerning the FHC by
establishing the connection of the partition function of the FHC with
a stochastic process described by a Fokker--Planck equation.
In the second part of Section~\ref{section:statistical}
the generating functional of the FJBC in the shish-kebab approximation
is derived. 
Next, in Section~\ref{section:four} we compute the generating
functional of the FJBC without any approximation.
In analogy with what
happens in the classical case, it turns out that this
generating functional is very
different from that of the FHC. However, we are able to show that they
both coincide in the limit of a continuous chain.
Finally, our conclusions are drawn in
Section~\ref{section:conclusions}.
\section{{ Classical dynamics}}\label{section:classical}

\subsection{The freely hinged chain (FHC)} \label{subsection:fhc}
\begin{figure}[h]
\centering
\includegraphics[height=8cm]{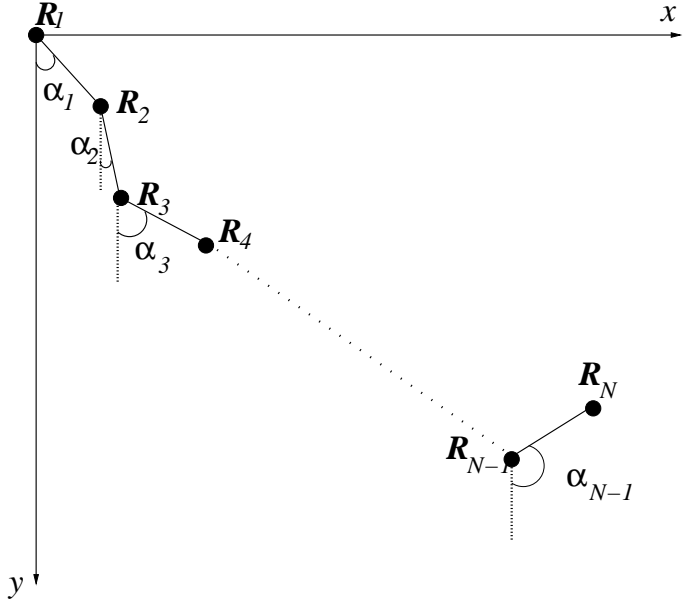}
\caption{FHC consisting of $N$ beads of mass $m$ (the black circles)
  connected together by $N-1$ segments of fixed length $a$. The angles
between the $y$ axis and the $i-$th segment are denoted by $\alpha_i$,
$i=1,\ldots,N-1$.}
\label{rysone}
\end{figure}

We begin by studying the dynamics of a chain
composed by $N$ beads of mass $m$ joined together by $N-1$ massless
segments of length $a$. Though it is not strictly necessary, to avoid
complications with the spherical coordinates in arbitrary dimensions,
in this Section the discussion will be limited 
to two dimensional chains moving on the 
$(x,y)$ plane. Let $\boldsymbol R_i(t)=(x_i(t),y_i(t))$ denote
the radius vector of the $i-$th bead, for $i=1,\ldots,N$. In polar
coordinates (see Fig.~\ref{rysone}):
\begin{eqnarray}
x_i(t)&=&\sum_{j=1}^{i-1}a\sin\alpha_{j}(t)\qquad\qquad
i=2,\ldots,N\label{xpolcoo}\\ 
y_i(t)&=&\sum_{j=1}^{i-1}a\cos\alpha_{j}(t)\qquad\qquad
i=2,\ldots,N\label{ypolcoo} 
\end{eqnarray}
We suppose that the first point $(i=1)$ is fixed in the origin, so
that:
\begin{equation}
x_1(t)=y_1(t)=0
\end{equation}
First of all, we compute the kinetic energy of the system:
\begin{equation}
T_{FHC}(a)=\sum_{i=2}^N\frac m2\left(
\dot{x}^2_i(t)+\dot{y}^2_i(t)
\right)\label{kinenedef}
\end{equation}
The dots denote here derivatives with respect to the time $t$,
i.~e. $\dot{x}= \frac{\partial x}{\partial t}$.
After a few calculations we find the expression of $T_{FHC}(a)$ in
polar coordinates:
\begin{equation}
T_{FHC}(a)=\frac{ma^2}2\sum_{i=2}^N
\sum_{j,k=1}^{i-1}\dot{\alpha}_j\dot{\alpha}_k
\cos(\alpha_j-\alpha_k)
\end{equation}
We are interested in the continuous chain obtained from the FHC after
performing the limit
in which the length $a$ of the segments goes to zero while the total
length $L$ of the chain is preserved:
\begin{equation}
a\longrightarrow 0 \qquad N\longrightarrow+\infty
\qquad\mbox{and}\qquad Na=L\label{contlim}
\end{equation}
In that limit the sums become integrals according to the prescription: 
\begin{equation}
\sum_{i=2}^{N}a\to\int_0^Lds\qquad\qquad
\sum_{j=1}^{i-1}a\to\int_0^sdu\label{rere} 
\end{equation}
A function $f_i(t)$ depending on the discrete  index $i$, like for
instance the angles 
$\alpha_i(t)$, will be substituted by a function of the continuous
variable $s$, where
$0\le s\le L$, according to the rule: $f_i(t)\longrightarrow
f(t,s)$.
The full procedure to
pass to the continuous limit (\ref{contlim}) is described in
Ref.~\cite{FePaVi,kleinert1} and will not be discussed here.
The new parameter $s$ appearing in (\ref{rere}) is the
arc--length that measures the 
distance between two points on the chain. Partial derivatives with
respect to $s$ will be denoted with a prime, for instance $\boldsymbol
R'(t,s)=\frac{\partial \boldsymbol R(t,s)}{\partial s}$.
We also need to express the mass of the beads $m$ as a function of
$a$. To this purpose, we introduce the density of mass $\rho$  on the
chain. Since the mass distribution is uniform -- all beads have the
same mass -- we may put  $\rho=\frac ML$, where $M$ is the total mass
of the chain. 
Thus,  $m$ can be written as a function of the mass density:
\begin{equation}
m=\frac ML a
\end{equation}
It is easy to check that,
for consistency, $Nm=M$ because
$Na=L$. 
Applying the above prescriptions, it is possible to show that the
continuous kinetic energy $T_{FHC}=\lim\limits_{a\to
  0,N\to+\infty,Na=L}T_{FHC}(a)$ is given by: 
\begin{equation}
T_{FHC}=\frac M{2L}\int_0^Lds\int_0^s du\int_0^sdv
{\dot\alpha}(t,u){\dot\alpha}(t,v)\cos\left(
\alpha(t,u)-\alpha(t,v)
\right)\label{kineticenergyfjccont}
\end{equation}
Let us note that the expression of $T_{FHC}$ given above is
different from that of Ref.~\cite{FePaVi}.
This is due to the fact that here we have not followed the recursive
procedure for computing $T_{FHC}(a)$ used in \cite{FePaVi}, but rather
we have derived it directly from the definition of kinetic energy of
Eq.~(\ref{kinenedef}).
It is possible to show the equivalence of the results obtained in this
work and in Ref.~\cite{FePaVi} exploiting the identity:%
\begin{equation}
\int_{0}^{s}du\int_{0}^{s}dvf(u,v)=\int_{0}^{s}du\int_{0}^{u}dvf(u,v)+\int
_{0}^{s}du\int_{0}^{u}dvf(v,u) \label{ideone}%
\end{equation}
which is valid for an arbitrary integrable function $f$ of $v$ and
$u$. Exploiting Eq.~(\ref{ideone}), the kinetic energy $T_{FHC}$ 
{\normalsize in (\ref{kineticenergyfjccont}) becomes:}%
\begin{equation}
T_{FHC}=\frac{M}{L}\int_{0}^{L}ds\int_{0}^{s}du\int_{0}^{u}dv\dot{\alpha
}(t,u)\dot{\alpha}(t,v)\cos(\alpha(t,u)-\alpha(t,v))
\end{equation}
The right hand side of the above equation coincides exactly with the
kinetic energy  
of the $ FHC$ {\normalsize derived in \cite{FePaVi}.}

Finally, we add to the FHC model the interactions. To this
purpose, we define the potential:
\begin{equation}
V_{FHC}(a)=\sum_{i=1}^NaV_E(\boldsymbol R_i(t)) +
\sum_{\underset{i\ne j}{i,j=1}}^Na^2 V_I(\boldsymbol
R_i(t),\boldsymbol R_j(t))\label{potclasfhc}
\end{equation}
where $V_E(\boldsymbol R_i(t))$ 
and $ V_I(\boldsymbol R_i(t),\boldsymbol R_j(t))$
have the meaning of energy densities per unit of chain length
due to  external and internal
forces respectively. For this reason, in the total potential energy of
the chain  $V_{FHC}(a)$ 
appearing in Eq.~(\ref{potclasfhc}),
$V_E(\boldsymbol R_i(t))$ 
and $ V_I(\boldsymbol R_i(t),\boldsymbol R_j(t))$ are multiplied by
the factors
$a$ and $a^2$ respectively.
The
presence of these factors is in agreement with the continuous limit
prescription of Eq.~(\ref{rere}).

In the continuous limit (\ref{contlim}) we obtain from $V_{FHC}(a)$:
\begin{equation}
{\normalsize V_{FHC}=}
\lim_{a\to0,N\to+\infty,Na=L} V_{FHC}(a)=
\int_{0}^{L}{\normalsize ds V}_{E}{\normalsize
  (\boldsymbol{R}% 
(t,s))+}\int_{0}^{L} ds\int_{0}^{L} ds' V_{I}{\normalsize (\boldsymbol{R}%
(t,s),\boldsymbol{R}(t,s}^{\prime}{\normalsize ))}
\label{potenecont}
\end{equation}
{\normalsize where }$\boldsymbol{R}(t,s)$ {\normalsize is the
  continuous limit 
of the bond vectors $\boldsymbol{R}_i(t)$.}
From Eqs.~(\ref{xpolcoo}) and (\ref{ypolcoo}) it turns out that the
components of
$\boldsymbol R(t,s)$ are given by:
\begin{equation}
\boldsymbol{R}{\normalsize (t,s)=}\int_{0}^{s}{\normalsize du(}\sin
{\normalsize \alpha(t,u),}\cos{\normalsize
  \alpha(t,u))}\label{contlimcoords} 
\end{equation}

\subsection{ The case of the FJBC}\label{subsection:2.1}

Let us consider a chain composed by ${\normalsize N}$ bars of
{\normalsize length }${\normalsize a}$. 
Each bar is regarded here as a one-dimensional segment with a uniform
distribution of mass along it.
The bars are joined at
the points:
\begin{equation}
x_{i}=\sum\limits_{i^{^{\prime}}=1}^{i-1}a\sin\alpha_{i^{^{\prime}}}.
\end{equation}%
\begin{equation}
y_{i}=\sum\limits_{i^{^{\prime}}=1}^{i-1}a\cos\alpha_{i^{^{\prime}}}.
\end{equation}
{\normalsize where $i = 2,
\ldots,N$. One end of the first bar is supposed to be fixed in the
origin, so that }$x_{1}=y_{1}=0.$ {\normalsize The coordinates of every point
of such a chain are given by:}%
\begin{equation}
x_{i}\left(  t,l\right)  =\sum\limits_{j=1}^{i-1}a\sin\alpha_{j}(t)+l\sin
\alpha_{i}(t)
\end{equation}%
\begin{equation}
y_{i}\left(  t,l\right)  =\sum\limits_{j=1}^{i-1}a\cos\alpha_{j}(t)+l\cos
\alpha_{i}(t)
\end{equation}
for $i=2,\ldots,N-1$.
The points of the first bar have
coordinates:%
\begin{equation}
{\normalsize x}_{i}\left(  t,l\right)  {\normalsize =l}\sin{\normalsize \alpha
}_{i}(t)%
\end{equation}%
\begin{equation}
y_{i}\left(  t,l\right)  =l\cos\alpha_{i}(t)%
\end{equation}
Here $l$ is a continuous parameter taking its values in the range
$0\le l\le a$.
{\normalsize In the future also the notation}
\begin{equation}
\boldsymbol{R}_{i}(t,l)=(x_{i}(t,l),y_{i}(t,l)),\qquad i=1,\ldots,N-1
\end{equation}
{\normalsize will be used. We are now ready to compute the kinetic
  energy of 
the } FJBC:
\begin{equation}
T_{FJBC}(a)=\sum_{i=1}^{N-1}\int_{0}^{a}\frac{\mu_i}{2a}dl\left(  \dot{x}_{i}%
^{2}(t,l)+\dot{y}_{i}^{2}(t,l)\right)
\end{equation}
{\normalsize In the above equation $\mu_i$ 
is the mass of the $i-$th bar. We have assumed that the mass
  distribution of each bar is 
homogeneous. For that reason, the mass }${\normalsize dm}_{i}$
{\normalsize of a small element of length }${\normalsize dl}$%
{\normalsize \ is:}%
\begin{equation}
dm_{i}=\frac{\mu_i}{a}dl
\end{equation}
If all bars have the same mass, i.~e. $\mu_i=\mu$ for
$i=1,\ldots,N-1$, it is possible to put $dm_i=\frac{M}{L}dl$ using the
fact that $\mu N=M$ and $aN=L$.
After simple calculations we find out that:%
\begin{equation}
T_{FJBC}(a)=\frac{M}{2L}\sum_{i=1}^{N-1}\int_{0}^{a}dl\left[  \sum
_{j,k=1}^{i-1}a^{2}\dot{\alpha}_{j}\dot{\alpha}_{k}\cos(\alpha_{j}-\alpha
_{k})+l^{2}\dot{\alpha}_{i}^{2}+2al\sum_{j=1}^{i-1}\dot{\alpha}_{j}\dot
{\alpha}_{i}\cos(\alpha_{j}-\alpha_{i})\right]
\end{equation}
{\normalsize After the integration over $dl$ we obtain:}%
\begin{equation}
T_{FJBC}(a)=\frac{M}{2L}\sum_{i=1}^{N-1}a^{3}\left[  \sum_{j,k=1}^{i-1}%
\dot{\alpha}_{j}\dot{\alpha}_{k}\cos(\alpha_{j}-\alpha_{k})+\frac{\dot{\alpha
}_{i}^{2}}{3}+\sum_{j=1}^{i-1}\dot{\alpha}_{j}\dot{\alpha}_{i}\cos(\alpha
_{j}-\alpha_{i})\right]\label{kinenefjbc}
\end{equation}
Let us note that, in order to derive the above expression
of the kinetic energy, the  computation of the
moments of inertia of the bars  has not been necessary. Each bar has
been rather decomposed
into a set of infinitesimal elements which can
be regarded as points. The kinetic energy
of each of these points has 
then been summed up to obtain the total kinetic energy of
the chain given in
Eq.~(\ref{kinenefjbc}). 
The result is the same as if we
had considered the bars as the basic
units of the chain and computed their kinetic energy.
An approach in which the chain is regarded as a set of points and not
as a system of bars is not a great advantage
in the simple classical case. However, it will be of
crucial importance when the statistical physics of the FJBC
will be studied, because the Fokker-Planck equation for a bar is
prohibitively complicated for our purposes.

{\normalsize We are now ready to perform the continuous limit
  (\ref{contlim}) 
in which
  the length $a$ of the segments goes to zero while the total length
  $L$ of 
  the chain is preserved.
Putting}%
\begin{equation}
\lim_{a\to0,N\to+\infty,Na=L}T_{FJBC}(a)=T_{FJBC}%
\end{equation}
{\normalsize the continuous kinetic energy takes the form:}%
\begin{equation}
T_{FJBC}=\frac{M}{2L}\int_{0}^{L}ds\int_{0}^{s}du\int_{0}^{s}dv\dot{\alpha
}(t,u)\dot{\alpha}(t,v)\cos(\alpha(t,u)-\alpha(t,v)) \label{tfjbc}%
\end{equation}
Eq.~(\ref{tfjbc}) coincides exactly with the kinetic energy of the
continuous FHC of Eq.~(\ref{kineticenergyfjccont}).

{\normalsize Let's now add to the }$FJBC$ the {\normalsize interactions
described by a general potential of the kind:}%
\begin{equation}
{\normalsize V_{FJBC}(a)=}\sum_{i=1}^{N-1}\int_{0}^{a}{\normalsize
  dl\tilde{V}}_{E}% 
{\normalsize (}\boldsymbol{R}_{i}{\normalsize (t,l))+}\sum
_{\substack{i,j=1\\i\neq j}}^{N-1}\int_{0}^{a}{\normalsize dl}\int_{0}%
^{a}{\normalsize dl}^{\prime}{\normalsize \tilde{V}}_{I}{\normalsize (}\boldsymbol{R}%
_{i}{\normalsize (t,l),}\boldsymbol{R}_{j}{\normalsize (t,l}^{\prime
}{\normalsize ))}\label{potdensfjbccont}%
\end{equation}
{\normalsize Here }$\tilde{V}_{E}(\boldsymbol{R}_{i})$ {\normalsize and }%
$\tilde{V}_{I}(\boldsymbol{R}_{i},\boldsymbol{R}_{j})$ {\normalsize
  take into account 
the external and internal interactions respectively. In the case of small
values of }$a${\normalsize , }$V_{FJBC}(a)$ {\normalsize may be
  approximated as 
follows:}%
\begin{equation}
{\normalsize V_{FJBC}(a)\sim}\sum_{i=1}^{N-1}{\normalsize
  a\tilde{V}}_{E}{\normalsize (}% 
\boldsymbol{R}_{i}{\normalsize (t,a))+}\sum_{\substack{i,j=1\\i\neq j}%
}^{N-1}{\normalsize a}^{2}{\normalsize \tilde{V}}_{I}{\normalsize
  (}\boldsymbol{R}% 
_{i}{\normalsize (t,a),}\boldsymbol{R}_{j}{\normalsize
  (t,a))}\label{ascaleemerg} 
\end{equation}
{\normalsize with}%
\begin{equation}
\boldsymbol{R}_{i}{\normalsize (t,a)=}\sum_{j=1}^{i-1}{\normalsize a(}%
\sin{\normalsize \alpha}_{j}{\normalsize (t),}\cos{\normalsize \alpha}%
_{j}{\normalsize (t))}\qquad\qquad i=2,\ldots,N \label{bond vector}%
\end{equation}
and $\boldsymbol{ R}_1(t)=0$.
{\normalsize The above approximation of }$V_{FJBC}(a)$ {\normalsize
  becomes exact in the 
continuous limit of Eq.~(\ref{contlim}), in which the expression of the
potential becomes:}%
\begin{equation}
{\normalsize V_{FJBC}=}\int_{0}^{L}{\normalsize ds \tilde{V}}_{E}{\normalsize
  (\boldsymbol{R}% 
(t,s))+}\int_{0}^{L}{\normalsize ds
\int_0^Lds'
\tilde{V}_I}{\normalsize (\boldsymbol{R}%
(t,s),\boldsymbol{R}(t,s}^{\prime}{\normalsize ))}%
\end{equation}
{\normalsize where }$\boldsymbol{R}(t,s)$ is given in
  Eq.~(\ref{contlimcoords}). As we may see, the above expression of the
  potential coincides with that of the FHC of Eq.~(\ref{potenecont}).
As a consequence, in the continuous limit the FHC and FJBC are
equivalent as expected, despite the fact that the discrete models have
different kinetic and potential energies.
%  continuous limit 
%of the bond vectors given in Eq. (\ref{bond vector}):}%
%\begin{equation}
%\boldsymbol{R}{\normalsize (t,s)=}\int_{0}^{s}{\normalsize du(}\sin
%{\normalsize \alpha(t,u),}\cos{\normalsize \alpha(t,u))}%
%\end{equation}

\section{Chain statistical dynamics}\label{section:statistical}

\subsection{{\protect\normalsize The case of the }FHC}\label{fhccase}

{\normalsize In this section we consider the motion of a FHC 
composed by }$N$ {\normalsize beads in }$d$ {\normalsize
  dimensions. Each bead 
is fluctuacting in a viscous medium kept of constant temperature }$T$.
{\normalsize The viscosity makes the motion overdamped \cite{teraoka},
  so that the 
  radius vectors of the beads satisfy
 the Langevin equation:}%
\begin{equation}
\boldsymbol{\dot{R}}_{i}=\boldsymbol{\nu}_{i}{\normalsize (t),}\text{\qquad
}{\normalsize i=1,\ldots,N.}\label{langefree}%
\end{equation}
where $\boldsymbol{\nu}_{i}(t)$ {\normalsize is a random force with a
probability
distribution that will be specified later.}

%\begin{equation}
%\mathcal{D}{\normalsize \rho(}\boldsymbol{\nu}_{i}{\normalsize )=}%
%\mathcal{D}\boldsymbol{\nu}_{i}{\normalsize (t)}\exp\left[  -\frac{1}{4D}%
%\int_{0}^{t_{f}}\boldsymbol{\nu}_{i}^{2}(t)dt\right]
%\end{equation}
%${\normalsize D}$ {\normalsize being the diffusion constant of the
%beads. 
Due to the constraints, the above equations
have to be completed by conditions:
\begin{equation}
C_{i}(\boldsymbol{R}_{i},\boldsymbol{R}_{i-1})=\left\vert \boldsymbol{R}%
_{i}(t)-\boldsymbol{R}_{i-1}(t)\right\vert -{\normalsize a=0,\qquad
}i=2,\ldots,N.\label{constrequ}%
\end{equation}
It is easy to solve Eq. (\ref{langefree}) with respect to
the $\boldsymbol{R}_{i}$'s. The result is:% 
\begin{equation}
\boldsymbol{R}_{i,\boldsymbol{\nu}_{i}}(t)=\int_{0}^{t}\boldsymbol{\nu
}_{i}(\tau)d\tau+\boldsymbol{R}_{0,i}\label{solform}%
\end{equation}
We have denoted the solutions of Eq.~(\ref{langefree}) with the symbol
$\boldsymbol{R}_{i,\boldsymbol{\nu}_{i}}(t)$. The superscript
$\boldsymbol{\nu}_{i}$
is used to stress the dependence of these solutions on the noise.
Of course, not all components of the noise are
independent. 
In fact, if we plug in the solutions
$\boldsymbol{R}_{i,\boldsymbol{\nu}_i}(t)$'s of Eq.~(\ref{langefree}) in
Eqs.~(\ref{constrequ}), we obtain
constraints for the noises
$\boldsymbol{\nu}_{i}$'s:
\begin{equation}
C_{i}(\boldsymbol{R}_{i,\boldsymbol\nu_i},
\boldsymbol{R}_{i-1,\boldsymbol\nu_{i-1}})=
\left\vert
\boldsymbol{R}_{i,\boldsymbol\nu_i}(t)
-\boldsymbol{R}_{i-1,\boldsymbol\nu_{i-1}}(t)
\right\vert 
-{\normalsize a=0,\qquad
}i=2,\ldots,N.\label{constrequnu}%
\end{equation}
The system of equations (\ref{constrequnu})
allows in principle to eliminate the  degrees of
freedom of the random forces $\boldsymbol\nu_i$ that become spurious
because the beads 
are fixed at the ends of segments of constant length $a$.
Unfortunately, even in the present simple case in which the external
forces are absent, it is not easy to solve
Eqs.~(\ref{constrequnu}) by expressing the redundant $N-1$
 degrees of freedom as a function of the
remaining $2N+1$ independent variables.

An alternative procedure
consists in the introduction of a set of Lagrange multipliers
 $\lambda_{i}(t),$ $i=2,\ldots,N$ in order to impose the
constraints. In this approach the $\boldsymbol{\nu}_{i}$'s are
regarded as unconstrained sources of white noise. The price to be paid for
that is the addition of the reaction forces
\begin{equation}
\boldsymbol F_{R,i}=\lambda_i\frac{\partial
  C_i(\boldsymbol{R}_{i},\boldsymbol{R}_{i-1})}{\partial \boldsymbol
  R_i} \qquad\qquad i=2,\ldots,N
\end{equation}
in the 
Langevin equations (\ref{langefree}).
The solutions $\boldsymbol{R}_{i,\boldsymbol\nu_i,\lambda_i}$ of the
new Langevin
equations obtained in this way
will depend
also on the Lagrange multipliers $\lambda_i$. The latter can be
determined by exploiting the constraints (\ref{constrequ}).
In practice, it is more convenient to use the consistency
relations 
coming from the requirement that the constraints should be preserved
in time:

%In practice, it is not convenient to fix the 
%$\boldsymbol{\nu}_{i}$'s using the constraints themselves, it is much
%better to exploit the consistency requirements that the constraints
%are preserved in time, i.~e.:
\begin{equation}
\dot{C}%
_{i}(\boldsymbol{R}_{i,\boldsymbol{\nu}_{i},\lambda_i},\boldsymbol{R}%
_{i-1,\boldsymbol{\nu}_{i-1},\lambda_{i-1}})=0\label{rrr}
\end{equation}
%Let us denote with $\boldsymbol{R}^{sol}_{i,\boldsymbol{\nu}_i}$ the
%solutions of Eq.~(\ref{langefree}) that satisfy additionally
%Eqs.~(\ref{rrr}).
Explicitly, we have that
\begin{equation}
\dot{C}_{i}(\boldsymbol{R}_{i,\boldsymbol{\nu}_{i},\lambda_i},
\boldsymbol{R}_{i-1,\boldsymbol{\nu}_{i-1}\lambda_i})=\frac{1}{a}
(\boldsymbol{R}_{i,\boldsymbol{\nu}_{i},\lambda_i}-
\boldsymbol{R}_{i-1,\boldsymbol{\nu}_{i-1},\lambda_i})
\cdot(\boldsymbol{\nu}_{i}-\boldsymbol{\nu}_{i-1})=0\label{constrnu}
\end{equation}
for $i=2,\ldots,N$. 
%As a consequence of the above relations, it turns
%out that only a number 
%$N(d-1)+1$
%of the initial $Nd$ degrees of freedom of
%the noises $\boldsymbol{\nu}_{i}$ is
%independent. 
%Eqs. (\ref{constrnu}) are quite complicated to be 
%solved analytically. For this reason, 
%instead of restricting oneself to the set of independent degrees of
%freedom solving the constraints or Eqs.~\ref{constrnu} directly, it is
%preferable to introduce a set of Lagrange
%multipliers $\lambda_{i}(t),$ $i=2,\ldots,N$ in order to impose the
%constraints (\ref{constrequ}). In this approach one should add 
%to the
%Langevin equation (\ref{langefree})
%the reaction
%forces $\lambda_{i}\frac{\partial
%  C_{i}}{\partial\boldsymbol{R}_{i}}$. 
The constraints fixed with the help of the
Lagrange multipliers are sometimes 
called rigid constraints. 
The resulting Langevin equations are
mathematically complicated, but yet  their
solutions may 
be derived numerically.

With both methods, direct solution of Eqs.~(\ref{constrequnu}) or the
introduction of the Lagrange multipliers, 
it is very difficult to arrive to
a path
integral formulation  of an inextensible chain that can be used to
perform analytical calculations.
We recall that we are interested here in the construction of the
so-called generating functional
of the correlation
functions
of the solutions of the Langevin equations
 $\boldsymbol{R}_{i,\boldsymbol{\nu}_{i}}^{{}}$'$s$. 
It will be shown in this Section that it is possible to achieve this
goal starting from a slightly different point of view. 
The idea is to replace the usual gaussian noise distribution by 
%that strongly suppresses the spurious degrees of freedom and
%completely eliminates them in some suitable limit of the parameters
%appearing in that term.
%To this purpose, we will choose for the noises
%$\boldsymbol \nu_i$'s
the probability
distribution ${\cal D}\rho(\{\boldsymbol \nu_i\})$ given by:
\begin{equation}
{\cal D}\rho(\{\boldsymbol \nu_i\})=\left[
\prod_{i=1}^N
{\cal D}\boldsymbol\nu_i(t)\right]
\exp\left[
-\int_0^{t_f}dt\left(\sum_{i=1}^N
\frac{\boldsymbol\nu_i^2(t)}{4D}+V(\boldsymbol
R_{1,\boldsymbol\nu_1},\ldots,\boldsymbol
R_{N,\boldsymbol\nu_N},\beta_2,\ldots,\beta_N) 
\right)
\right]\label{noisedistribution}
\end{equation}
The potential $V(\boldsymbol
R_{1,\boldsymbol\nu_1},\ldots,\boldsymbol
R_{N,\boldsymbol\nu_N},\beta_2,\ldots,\beta_N)$ will be selected 
according to the following criteria:
\begin{enumerate}
\item \label{criter1} When $\beta_2=\beta_3=\cdots=\beta_N=0$, the system is
unconstrained and the $\boldsymbol \nu_i$'s become purely gaussian
noises. 
\item \label{criter2} For
large values of the $\beta_i$'s, the potential $V(\boldsymbol
R_{1,\boldsymbol\nu_1},\ldots,\boldsymbol
R_{N,\boldsymbol\nu_N},\beta_2,\ldots,\beta_N)$ should exhibit a sharp
minimum near the region of noise configurations in which the
constraints (\ref{constrequnu}) are satisfied. 
\item \label{criter3} Finally, in the limit
$\beta_i=+\infty$, the potential   $V(\boldsymbol
R_{1,\boldsymbol\nu_1},\ldots,\boldsymbol
R_{N,\boldsymbol\nu_N},\beta_2,\ldots,\beta_N)$ should be infinite if
Eqs.~(\ref{constrequnu}) are not fulfilled and zero otherwise. 
\end{enumerate}
Clearly, thanks to the third condition all noise configurations that
do not conform with 
Eqs.~(\ref{constrequnu}) are eliminated. Only those for which the
constraints (\ref{constrequnu}) are satisfied remain.
%As a consequence, in the limit
%$\beta\longrightarrow+\infty$, it is expected that the springs will
%behave as rigid massless bars whose lengths coincide with the rest
%length $a$ of the springs. 

To derive a potential with the characteristics described above we can
use the physical intuition according to which a chain made of beads
and springs will 
behave as a chain of beads and links of fixed length in the limit of
infinitely large 
spring elastic 
constants.
Thus, we consider here a system of $N$ beads connected together by
$N-1$ springs. Let $\beta_i$, $i=2,\ldots,N$ be the elastic constant
of the  $i-$th spring and $a$ its rest length.
The elastic forces:
\begin{equation}
\boldsymbol F_i=\beta_i\left(
\left|
\boldsymbol R_i-\boldsymbol R_{i-1}
\right|-a
\right)\frac{(\boldsymbol R_i-\boldsymbol R_{i-1})}{|\boldsymbol
  R_i-\boldsymbol R_{i-1}|} - \beta_i\left(
\left|
\boldsymbol R_i-\boldsymbol R_{i+1}
\right|-a
\right)\frac{(\boldsymbol R_i-\boldsymbol R_{i+1})}{|\boldsymbol
  R_i-\boldsymbol R_{i+1}|}\qquad(i=2,\ldots,N-1)\label{fi2n-1}
\end{equation}
are acting on the internal beads with indexes $i=2,\ldots,N-1$.
The following forces
\begin{equation}
\boldsymbol F_1=\beta_2\left(
\left|
\boldsymbol R_1-\boldsymbol R_{2}
\right|-a
\right)\frac{(\boldsymbol R_1-\boldsymbol R_{2})}{|\boldsymbol
  R_1-\boldsymbol R_{2}|}\label{fone}
\end{equation}
\begin{equation}
\boldsymbol F_N=\beta_N\left(
\left|
\boldsymbol R_N-\boldsymbol R_{N-1}
\right|-a
\right)\frac{(\boldsymbol R_N-\boldsymbol R_{N-1})}{|\boldsymbol
  R_N-\boldsymbol R_{N-1}|}\label{ftwo}
\end{equation}
are applied instead on the beads lying at the ends of the chain.
While from the point of view of the mathematical complexity there is
no problem in considering $N-1$ independent elastic constants $\beta_i$,
in practice this is an unnecessary complication. For this reason, from
now on we will assume that the $\beta_i$'s are all equal:
\begin{equation}
\beta_i=\beta\qquad\qquad i=2,\ldots,N
\end{equation}
After this simplification, the potential corresponding to the
interactions 
(\ref{fi2n-1}--\ref{ftwo}) may be written as follows:
\begin{equation}
V(\boldsymbol R_1,\ldots,\boldsymbol
R_N,\beta)=\sum_{j=2}^NaV_j(|\boldsymbol
R_j-\boldsymbol R_{j-1}|,\beta)\label{pottot} 
\end{equation}
where
\begin{equation}
V_j(|\boldsymbol R_j-\boldsymbol R_{j-1}|,\beta)=\frac\beta 2\left(
|\boldsymbol R_j-\boldsymbol R_{j-1}|-a
\right)^2\label{potsing}
\end{equation}
The minima of the potentials (\ref{potsing}) and thus the minimum of
$V(\boldsymbol R_1,\ldots,\boldsymbol R_N,\beta)$ occur when the
conditions 
$|\boldsymbol R_j-\boldsymbol R_{j-1}|-a=0$, $j=2,\ldots,N$, are
verified. These are exactly the inextensible constraints of
Eqs.~(\ref{constrequ}). It is easy to show that $V(\boldsymbol
R_1,\ldots,\boldsymbol R_N,\beta)$ satisfies requirements
1.~--~3.

We are now ready to write down the expression of the
generating functional of the FHC, which is defined as the average of
the quantity $\exp\left[  -\sum_{i=1}^{N}a\int
_{0}^{t_{f}}\boldsymbol{J}_{i}(t)\cdot\boldsymbol{R}_{i,\boldsymbol{\nu}_{i}%
}(t)dt\right]$ with respect to the probability distribution  ${\cal
  D}\rho(\{\boldsymbol \nu_i\})$ of Eq.~(\ref{noisedistribution}):
\begin{align}
Z_{FHC}[J]  & =\lim_{\beta\to+\infty}\left[
  \prod_{i=1}^{N}\int\mathcal{D}\boldsymbol{\nu}% 
_{i}(t)\right]  \exp\left[  -\int_{0}^{t_{f}}dt
\left(\sum_{i=1}^{N}
\frac{\boldsymbol{\nu}_{i}^{2}(t)}{4D} 
+\sum_{j=2}^{N}aV_{j}\left(  \left\vert
\boldsymbol{R}_{j,\boldsymbol{\nu}_{j}}^{{}}
-\boldsymbol{R}_{j-1,\boldsymbol{\nu}_{j-1}}^{{}}\right\vert,\beta
\right)  \right)\right]
  \nonumber
\\
&  \exp\left[  -\sum_{i=1}^{N}a\int
_{0}^{t_{f}}\boldsymbol{J}_{i}(t)\cdot\boldsymbol{R}_{i,\boldsymbol{\nu}_{i}%
}(t)dt\right]
\label{zfjcdef} 
\end{align}
Let us note that in Eq.~(\ref{zfjcdef}) the
%$\boldsymbol{R}_{i,\boldsymbol{\nu}_{i}}$'$s$ coincide with the
%solutions of the Langevin equations 
%given in Eq. (\ref{solform}), thus before imposing the constraints.
path integration is extended over all possible noise configurations. 
As already anticipated before,
the
constraints are fixed using the potentials
$V_j\left(  \left\vert
\boldsymbol{R}_{j,\boldsymbol{\nu}_j}-\boldsymbol{R}_{j-1,\boldsymbol{\nu}_{j-1}}
\right\vert,\beta \right)$ after taking the limit
$\beta\longrightarrow+\infty$. Indeed, when
$\beta$ is 
very large
all configurations $\boldsymbol{R}_{i,\boldsymbol{\nu}_{i}}$ that do not satisfy
the conditions $C_{i}(\boldsymbol{R}_{i,\boldsymbol{\nu}_{i}},\boldsymbol{R}%
_{i-1,\boldsymbol{\nu}_{i-1}})=0$ are exponentially
suppressed. Eq.~(\ref{zfjcdef}) should be completed by specifying
the positions $\boldsymbol{R}_{i,\boldsymbol{\nu}_{i}}$ of the beads
at the initial and final instants. 
We require that
$\boldsymbol{R}_{i,\boldsymbol{\nu}_{i}}(0)=\boldsymbol R_{0,i}$
according to Eq.~(\ref{solform}). Moreover, 
when
$t=t_f$ 
it is assumed that the $i-$th bead is located at the point $\boldsymbol
R_{f,i}$. Of course, 
both $\boldsymbol R_{0,i}$ and $\boldsymbol
R_{f,i}$ should satisfy the constraints (\ref{constrequ}).

At this point we
insert in Eq. (\ref{zfjcdef}) the quantity%
%\begin{align}
%Z_{FHC}[J]  & =\left[  \prod_{i=1}^{N}\int\mathcal{D}\boldsymbol{\nu}%
%_{i}(t)\right]  \exd p\left[  -\frac{1}{4D}\sum_{i=1}^{N}\int_{0}^{t_{f}%
%}\boldsymbol{\nu}_{i}^{2}(t)dt\right]  \exp\left[  -\sum_{i=1}^{N}\int
%_{0}^{t_{f}}\boldsymbol{J}_{i}(t)\cdot\boldsymbol{R}_{i,\boldsymbol{\nu}_{i}%
%}(t)dt\right]  \nonumber
%\\
%& \exp\left[  -\sum_{j=2}^{N}\int_{0}^{t_{f}}V_{j}\left(  \left\vert
%\boldsymbol{R}_{j,\boldsymbol{\nu}_{j}}^{{}}-\boldsymbol{R}%
%_{j-1,\boldsymbol{\nu}_{j-1}}^{{}}\right\vert \right)  dt\right]\label{zfjcdef}
%\end{align}
\begin{equation}
I=\prod\limits_{i=1}^{N}\int_{\boldsymbol
  r_{i}(0)=\boldsymbol R_{0,i}}^{\boldsymbol r_i(t_f)=\boldsymbol R_{f,i}}
\mathcal{D}\boldsymbol{r}_{i}\delta(\boldsymbol{r}%
_{i}-\boldsymbol{R}_{i,\boldsymbol{\nu}_{i}})\label{step1}
\end{equation}
The boundary conditions of the integration over the new fields
$\boldsymbol r_i(t)$ have been chosen in such a way that they are
consistent with the boundary conditions of the fields $\boldsymbol
R_{i,\boldsymbol \nu_i}$.
Clearly $I=1$, so that the insertion of $I$ 
in Eq.~(\ref{zfjcdef}) will not change the
physics of the 
problem. As a consequence, 
remembering the explicit expressions of the potentials $V_i(|\boldsymbol{R}_i-
\boldsymbol{R}_{i-1}|,\beta)$ of Eq.~(\ref{potsing}),
we may rewrite $Z_{FHC}[J]$ as follows:
\begin{align}
Z_{FHC}[J] &  =\lim_{\beta\to+\infty}\left[
  \prod_{i=1}^{N}\int\mathcal{D}\boldsymbol{\nu}_{i}
\int_{\boldsymbol r_i(0)=\boldsymbol R_{0,i}}^{\boldsymbol
  r_i(t_f)=\boldsymbol R_{f,i}}
\mathcal{D}\boldsymbol{r}_{i}\right]  \exp\left[  -\frac{1}{4D}\sum
_{i=1}^{N}\int_{0}^{t_{f}}\boldsymbol{\nu}_{i}^{2}(t)dt\right]  \left[
\prod\limits_{i=1}^{N}\delta(\boldsymbol{r}_{i}-\boldsymbol{R}%
_{i,\boldsymbol{\nu}_{i}})\right] \nonumber \\
&  \exp\left[  -\sum_{i=1}^{N}a\int_{0}^{t_{f}}\boldsymbol{J}_{i}(t)%
\cdot\boldsymbol{r}_{i}(t)dt\right]  \exp\left[  -a\beta\sum_{j=2}^{N}\int
_{0}^{t_{f}}\left(  \left\vert \boldsymbol{r}_{j}-\boldsymbol{r}_{j-1}
\right\vert-a\right)^{2}dt\right]
\end{align}
Now we recall the identity:
\begin{equation}
\delta(\boldsymbol{r}_{i}-\boldsymbol{R}_{i,\boldsymbol{\nu}_{i}})=\det\left[
\frac{\partial}{\partial t}\right]  \delta(\boldsymbol{\dot{r}}_{i}%
-\boldsymbol{\nu}_{i})\label{step2}
\end{equation}
where we have used the fact that $\boldsymbol{R}_{i,\boldsymbol{\nu}_{i}}$
satisfies Eq. (\ref{langefree}). Ignoring the constant factor $\prod
\limits_{i=1}^{N}\det\left[  \frac{\partial}{\partial t}\right]  $ and
performing the simple integrations over the noises $\boldsymbol{\nu}_{i}$, the
generating functional $Z_{FHC}[J]$ takes the form:
\begin{eqnarray}
Z_{FHC}[J]&=&\lim_{\beta\to+\infty}\left[
\prod_{i=1}^N\int_{\boldsymbol r_i(0)=\boldsymbol R_{0,i}}^{\boldsymbol
  r_i(t_f)=\boldsymbol R_{f,i}}{\cal D}\boldsymbol r_i(t)
\right]\exp\left[
-\sum_{i=1}^N\int_0^{t_f}\left(\frac{1}{4D}\dot{\boldsymbol r}_i^2
+a\boldsymbol J_i\cdot\boldsymbol r_i
\right)dt
\right]\nonumber\\
&\times&\exp \left[-a\beta\sum_{j=2}^N\int_0^{t_f}\left(
|\boldsymbol r_j-\boldsymbol r_{j-1}|
-a\right)^2dt\right]\label{zfjcabc}
\end{eqnarray}
Before continuing, a digression is in order.
First, we rewrite $Z_{FHC}[J]$ as follows:
\begin{equation}
Z_{FHC}[J]=\lim_{\beta\to+\infty}\int
{\cal
  D}\rho_\beta(\{\boldsymbol r_i(t)\})e^{-\sum_{i=1}^N\int_0^{t_f}a\boldsymbol
  J_i\cdot\boldsymbol r_i} \label{fhcconstrpf}
\end{equation}
where 
${\cal
  D}\rho_\beta(\{\boldsymbol r_i(t)\})$ is the probability distribution:
\begin{eqnarray}
{\cal
  D}\rho_\beta(\{\boldsymbol r_i(t)\})&=&\left[
\prod_{i=1}^N{\cal D}
\boldsymbol r_i(t)
\right]\exp\left[
-\int_0^{t_f}dt\left(
\sum_{i=1}^N\frac{\dot{\boldsymbol r}_i^2}{4D}+V(\boldsymbol
r_1,\ldots,\boldsymbol r_N,\beta) 
\right)
\right]\nonumber\\
&\times&\prod_{i=1}^N\left[\delta(\boldsymbol
  r_i(t_f)-\boldsymbol R_{f,i}) 
\delta(\boldsymbol r_i(0)-\boldsymbol R_{0,i})\right]
\end{eqnarray}
The last two delta functions are needed to impose the boundary conditions.
For each fixed value of $\beta$ the probability function
$P_\beta(\{R_{f,i}\},t_f;\{R_{0,i}\},0)$ of the system of beads and
springs is defined
as the integral
over all possible configurations of the above probability distribution
\cite{Tanimura}:
\begin{equation}
P_\beta(\{R_{f,i}\},t_f;\{R_{0,i}\},0)=
\left[
\prod_{i=1}^N\int_{\boldsymbol r_i(0)=\boldsymbol
  R_{0,i}}^{\boldsymbol r_i(t_f)=\boldsymbol R_{f,i}} {\cal D}\boldsymbol r_i(t) 
\right]\exp\left[
-\int_0^{t_f}dt\left(
\sum_{i=1}^N\frac{\dot{\boldsymbol r}_i^2}{4D}+V(\boldsymbol
r_1,\ldots,\boldsymbol r_N,\beta) 
\right)
\right]\label{partfuncdef}
\end{equation}
The probability function
$P(\{R_{f,i}\},t_f;\{R_{0,i}\},0)$  of the FHC is obtained in the 
limit $\beta\to+\infty$:
\begin{equation}
P(\{R_{f,i}\},t_f;\{R_{0,i}\},0)=\lim_{\beta\to+\infty}
P_\beta(\{R_{f,i}\},t_f;\{R_{0,i}\},0)
\end{equation}
Let us note that $P_\beta(\{R_{f,i}\},t_f;\{R_{0,i}\},0)$
 satisfies the Schr\"odinger-like equation of a system of particles
 with interactions described by the  elastic potential (\ref{pottot}):
\begin{equation}
\left[
\frac{\partial}{\partial t_f}-\frac
12D\sum_{i=1}^N\frac{\partial^2}{\partial \boldsymbol
  R_{f,i}^2}+V(\boldsymbol R_{f,1},\ldots,\boldsymbol R_{t_f,n},\beta)
\right]P_\beta(\{R_{f,i}\},t_f;\{R_{0,i}\},0)=0\label{schroed-like}
\end{equation}
As it is possible to see, 
the elastic interactions do not appear inside a drift term as it
happens 
in standard Fokker-Planck equations. 
This is not a surprise, because
the elastic forces do not describe any physical property of the
chain. They have been introduced in the non-gaussian
noise distribution of 
Eq.~(\ref{noisedistribution}) with the sole purpose of imposing the
constraints.
On the other side, the generating functional
$Z_{FHC}[J]$ is related to a stochastic equation.
To show that this is exactly the case, we use the well known
connection between
Schr\"odinger--like and Fokker--Planck equations.
First of all, we introduce the new potential $U(\boldsymbol
R_{f,1},\ldots,\boldsymbol R_{f,N},\beta)$ through the differential
equation:
\begin{equation}
\frac D2\sum_{i=1}^N\left[
\frac{\partial^2 U}{\partial \boldsymbol R_{f,i}^2}(\boldsymbol
R_{f,1},\ldots,\boldsymbol R_{f,N},\beta) +\left(
\frac{\partial U}{\partial \boldsymbol R_{f,i}}(\boldsymbol
R_{f,1},\ldots,\boldsymbol R_{f,N},\beta)
\right)^2
\right]=V(\boldsymbol
R_{f,1},\ldots,\boldsymbol R_{f,N},\beta)
\end{equation}
Let us note that  $U(\boldsymbol
R_{f,1},\ldots,\boldsymbol R_{f,N},\beta)$ is dimensionless, so that
we may rescale the partition function
$P_\beta(\{R_{f,i}\},t_f;\{R_{0,i}\},0)$ as follows:
\begin{equation}
P_\beta'(\{R_{f,i}\},t_f;\{R_{0,i}\},0)=P_\beta(\{R_{f,i}\},t_f;\{R_{0,i}\},0)
e^{U(\boldsymbol
R_{f,1},\ldots,\boldsymbol R_{f,N},\beta)}
\end{equation}
It is easy to check that the new partition function
$P_\beta'(\{R_{f,i}\},t_f;\{R_{0,i}\},0)$
satisfies the Fokker--Planck equation:
\begin{equation}
\left[
\frac{\partial}{\partial t_f}-\frac
12\sum_{i=1}^N\frac{\partial}{\partial \boldsymbol R_{f,i}}\left(
D\frac{\partial}{\partial \boldsymbol R_{f,i}}+\boldsymbol f_i(\boldsymbol
R_{f,1},\ldots,\boldsymbol R_{f,N},\beta)
\right)
\right]P_\beta'(\{R_{f,i}\},t_f;\{R_{0,i}\},0)=0\label{stochproc}
\end{equation}
where the forces $\boldsymbol f_i(\boldsymbol
R_{f,1},\ldots,\boldsymbol R_{f,N},\beta)$ are given by:
\begin{equation}
\boldsymbol f_i(\boldsymbol
R_{f,1},\ldots,\boldsymbol R_{f,N},\beta)=-2D\frac{
\partial U(\boldsymbol
R_{f,1},\ldots,\boldsymbol R_{f,N},\beta)}{\partial \boldsymbol
  R_{f,i}}
\label{forces}
\end{equation}
This concludes our proof.

Let's now go back to the
generating functional of Eq.~(\ref{zfjcabc}).
Its expression may be further simplified exploiting
 the relation:
\begin{equation}
\lim_{\beta\to+\infty}e^{-a\beta\int_0^{t_f}f^2(\boldsymbol r_i(t),\boldsymbol
  r_j(t))dt}=\delta(f(\boldsymbol r_i(t),\boldsymbol r_j(t)))
\label{delide}
\end{equation}
which is valid up to an (infinite) proportionality constant for a
generic function $f(\boldsymbol r_i,\boldsymbol r_j)$. Eq.~(\ref{delide}) will
be proved in Appendix A. Applying Eq.~(\ref{delide}) in
Eq.~(\ref{zfjcabc}) in the limit of large values of $\beta$, we obtain:
\begin{eqnarray}
Z_{FHC}[J]&=&\left[
\prod_{i=1}^N\int_{\boldsymbol r_i(0) =\boldsymbol
  R_{0,i}}^{\boldsymbol r_i(t_f)=\boldsymbol R_{f,i}}{\cal D}\boldsymbol r_i(t)
\right]\exp\left[
-\sum_{i=1}^N\int_0^{t_f}\left(\frac{1}{4D}\dot{\boldsymbol r}_i^2
+a\boldsymbol J_i\cdot\boldsymbol r_i
\right)dt
\right]\nonumber\\
&\times&
\left[
\prod_{j=2}^N\delta\left(
|\boldsymbol r_j-\boldsymbol r_{j-1}|-a
\right)
\right]\label{genfunfjcdiscfin}
\end{eqnarray}
The above generating functional coincides, apart from an irrelevant
constant, to the generating functional of the discrete chain composed
by beads and links discussed in Ref.~\cite{FePaVi}. To show that,
first we note that inside Eq.~(\ref{genfunfjcdiscfin}) it is possible
to replace the functional Dirac delta functions
$\delta\left(
|\boldsymbol r_j-\boldsymbol r_{j-1}|-a
\right)$  with
$\delta\left(
|\boldsymbol r_j-\boldsymbol r_{j-1}|^2-a^2
\right)$.
A rigorous proof of this fact has been provided in Ref.~\cite{FePaVi}.
Basically the proof consists in the generalization to functional delta
functions of the following delta function relation
\begin{equation}
\delta(x^2-a^2)=\frac 12\left(
\delta(x-a)+\delta(x+a)
\right)\label{deltafunidenonfun}
\end{equation}
and on the consideration that the set of chain conformations for which
$|\boldsymbol r_j-\boldsymbol r_{j-1}|+a=0$ is empty.
Using also the identity:
\begin{equation}
\delta\left(
|\boldsymbol r_j-\boldsymbol r_{j-1}|^2-a^2
\right)=\left[\prod_{j=2}^Na^{-2}\right]
\delta\left(\frac
{|\boldsymbol r_j-\boldsymbol r_{j-1}|^2}{a^2}-1
\right)\label{adfdhfj}
\end{equation}
and neglecting irrelevant constants
we may rewrite Eq.~(\ref{genfunfjcdiscfin})  as follows:
\begin{eqnarray}
Z_{FHC}[J]&=&\left[
\prod_{i=1}^N\int_{\boldsymbol r_i(0) =\boldsymbol
  R_{0,i}}^{\boldsymbol r_i(t_f)=\boldsymbol R_{f,i}}{\cal D}\boldsymbol r_i(t)
\right]\exp\left[
-\sum_{i=1}^N\int_0^{t_f}\left(\frac{1}{4D}\dot{\boldsymbol r}_i^2
+a\boldsymbol J_i\cdot\boldsymbol r_i
\right)dt
\right]\nonumber\\
&\times&\left[
\prod_{j=2}^N\delta\left(
\frac{|\boldsymbol r_j-\boldsymbol r_{j-1}|^2}{a^2}-1
\right)
\right]\label{genfunfjcdiscfinplus}
\end{eqnarray}
This is exactly the generating functional of the FHC derived in
Ref.~\cite{FePaVi}. The continuous limit of
 $Z_{FHC}[J]$ is known
and coincides with the partition function of the GNL$\sigma$M:
\begin{equation}
Z[J]=\int_{\boldsymbol r(0,s)=\boldsymbol R_0(s)}^{\boldsymbol
  r(t_f,s)=\boldsymbol R_f(s)}{\cal D}\boldsymbol r(t,s)e^{-\frac
  1{2k_BT\tau}\frac{M}{2L}\int_0^{t_f}dt\int_0^Lds \dot{\boldsymbol
    r}^2(t,s)}
e^{-\int_0^{t_f}dt\int_0^Lds \boldsymbol J(t,s)\cdot\boldsymbol
  r(t,s)}\delta\left(
\left|\frac{\partial\boldsymbol r(t,s)}{\partial s}\right|^2-1
\right)\label{genfunfjc}
\end{equation}
Here $k_B$ is  the Boltzmann constant, $T$ is the temperature and 
$\tau$ is the relaxation time of the beads composing the chain.
These quantities arise from the factor $\frac 1{4D}$ appearing in
Eq.~(\ref{genfunfjcdiscfinplus}) and can be derived following the
procedure explained in \cite{FePaVi} and repeated here for convenience.
First,  we
note that $D=\mu kT$, where $\mu$ is the mobility of the
beads. Remembering the fact that $\mu=\frac \tau m$, we get 
$D=\frac{kT\tau}{m}$ and thus $\frac 1{4D}=\frac 1{2kT\tau}\frac
m2$. Exploiting the fact that $m=\frac M La$, we obtain the desired
expression: 
\begin{equation}
\frac 1{4D}=\frac{1}{2kT\tau}\frac
      {M}{2L}a\label{14D} 
\end{equation}
$Z[J]$ describes the dynamics of a continuous chain of length $L$. 
The
inextensibility constraints are imposed by the functional delta
function
$\delta\left(
\left|\frac{\partial\boldsymbol r(t,s)}{\partial s}\right|^2-1
\right)$.
\subsection{The case of the FJBC}\label{subsection:FJBC}
\begin{figure}
\centering
\includegraphics[height=6cm]{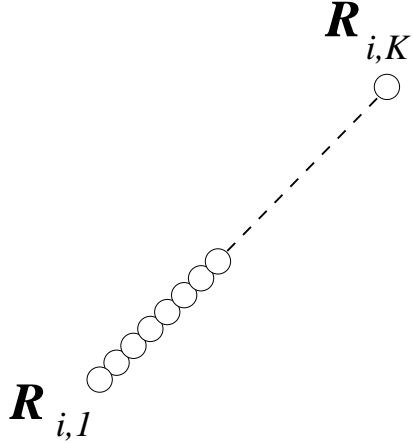}
\caption{Shish kebab model for the bars composing the chain.}
\label{figone} 
\end{figure}

With respect to Subsection~\ref{subsection:2.1}, which was dedicated
to the classical FJBC, in this Subsection 
we slightly change the definition of the small basic units composing
the FJBC. Instead of one-dimensional segments with uniform mass
distribution, each bar is replaced by 
a shish-kebab model
consisting of $K$ small spheres with diffusion constant
$D$ and mass $m$
 as shown in Fig.~\ref{figone}.
The distance $\Delta l$ of each sphere is given by:
\begin{equation}
\Delta l=\frac{a}{K}\label{distbetwspher}%
\end{equation}
In the case of dynamics it is possible to set the diameters of the
spheres by choosing their 
diffusion constants (or alternatively their mobility, their relaxation
times etc.) appropriately. Here we will assume that the diffusion
constant is given by $D=\frac{k_BT}{3\pi\eta\Delta l}$, where
$\eta$ is the viscosity of the medium in which the chain
fluctuates. This choice corresponds to spheres of diameter $\Delta
l$. It does not reduce the generality of our discussion, which remains
valid in the case of any other choice. 
The positions of the centers of mass of the $k-$th sphere belonging to
the $i-$th shish-kebab is described by the radius vectors
$\boldsymbol{R}_{i,k}$, $i=1,\ldots,N$ and $k=1,\ldots,K$. 
Throughout the
 rest of this Section the words shish-kebab and bar will be used
 interchangeably despite the fact that they are related to different objects.
In the next Section we will see that one-dimensional bars are
recovered in the limit $K\to+\infty$.  

At this point we are ready to impose the constraints that fix the
positions of the spheres in such a way that their motion will not
destroy the shape of the FJBC.
First of all, 
we need to derive the set of conditions satisfied by
the spheres belonging to
the same bar. 
To this purpose we note that if two spheres $k$ and
$k^{\prime}$ belong to the same $i-$th bar,
the distance between their centers of mass must be $\frac{a}
{K}|k-k^{\prime}|$. 
As a consequence, the radius vectors $\boldsymbol{R}_{i,k}$ and
$\boldsymbol{R}_{i,k'}$ have to
satisfy the relations:
\begin{equation}
\left\vert \boldsymbol{R}_{i,k^{\prime}}(t)-\boldsymbol{R}_{i,k}(t)\right\vert
=\frac{a}{K}|k-k^{\prime}|\qquad k\neq k^{\prime}\qquad k,k^{\prime
}=1,\ldots,K\label{consttwo}
\end{equation}
The above set of constraints is sufficient in order
to guarantee that the spheres in a bar will remain aligned during
their fluctuations.

\begin{figure}
\centering
\includegraphics[height=6cm]{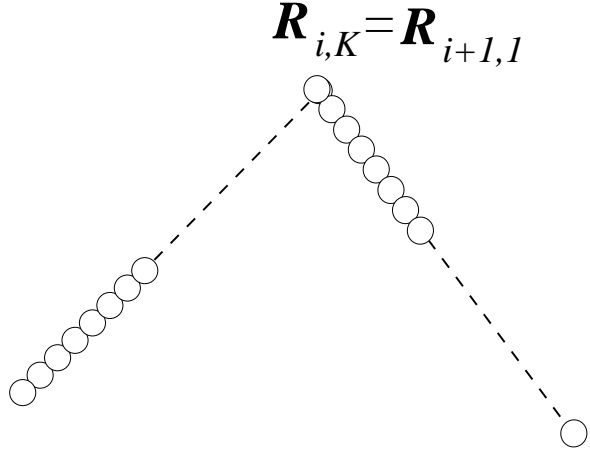}
\caption{This figure shows how the bars are connected together to
  build the FJBC.}
\label{figtwo}
\end{figure}

Next, we have to be sure that the ends of the bars are correctly connected
together in order to form
a chain. This goal is realized by
identifying the $K-$th sphere of the $i-$th bar  with
the first sphere of the 
$(i+1)-$th bar, see Fig. \ref{figtwo}.
This requirement implies that the locations of the centers of mass of
the spheres should be
constrained by the conditions:
\begin{equation}
\boldsymbol{R}_{i,K}=\boldsymbol{R}_{i+1,1}\qquad i=1,\ldots,N-1\label{constone}%
\end{equation}
Let us note that
$\boldsymbol{R}_{1,1}$ and $\boldsymbol{R}_{N,K}$ are not affected by the
constraints (\ref{constone}) because they are at the ends of the
chain which are free. 

Before continuing our main discussion, let's make a digression
concerning the length of the FJBC defined above.
The total length of the chain is not $L=Na$, because  we have
to subtract from this value
the length $\Delta l(N-1)$ of the $N$  spheres which are
identified
at the joints.
After doing that, we obtain the
effective length $\tilde L$ of the chain:
\begin{equation}
\tilde{L}=Na-\Delta l(N-1)
\end{equation}
Due to the fact that $\Delta l=\frac aK$, we get:
\begin{equation}
\tilde{L}=\frac{a(K-1)N}{K}+\frac{a}{K}\label{totlengthdischain}%
\end{equation}
%Of course, we should remember that the chain before the limit
%$a\longrightarrow 0$ is a three dimensional object and its length
%cannot be uniquely defined before taking the continuous limit, in
%which the singpheres become just points.
%The definition of the length $\tilde{L}$ given above corresponds to
%the case in which the length is measured along the lines connecting the
%centers of mass of the spheres, see Fig.~\ref{figthree}.
%\begin{figure}
%\centering
%\includegraphics[height=6cm]{lengthdef.eps}
%\caption{This figure shows the convention used in order to define the
%  length in Eq.~(\ref{totlengthdischain}). The length is measured
%  along the lines connecting the centers of mass of the spheres.}
%\label{figthree}
%\end{figure}

Let us now come back to the constraints (\ref{consttwo}) and (\ref{constone}).
They represent all the constraints that are needed to make sure that
the shape of the $FJBC$ is preserved during the motion of the spheres.
Luckily,
the constraints (\ref{constone}) may  be easily eliminated. This fact
was already noted in Ref.~\cite{liverpool} and is one of the main
advantages to construct an inextensible chains starting from a set of
rigid and inextensible basic units.
In order to get rid of the conditions (\ref{constone}), we choose as
independent coordinates 
the radius vectors $\boldsymbol{R}_{i+1,1}$'s. Accordingly,
the radius vectors $\boldsymbol R_{i,K}$'s have to be replaced
everywhere with the
$\boldsymbol{R}_{i+1,1}$'s. First of all,  we have to perform the
substitutions $\boldsymbol R_{i,K}\longrightarrow\boldsymbol{R}_{i+1,1} $
inside the relations
(\ref{consttwo}). In order to proceed,
it is convenient to divide these 
 relations into three sets.
The first set 
imposes conditions only on the coordinates
$\boldsymbol R_{i,k}$ for
$i=1,\ldots,N$ and $k=1,\ldots,K-1$.
These coordinates are not located at the junctions between two
neighboring bars and thus are
not affected 
by the elimination of the constraints
(\ref{constone}).
As a consequence,
in the restricted range of the indexes for which
$i=1,\ldots,N$ and $k\ne k'=1,\ldots,K-1$, we can simply
rewrite the constraints (\ref{consttwo}) without changes:
\begin{equation}
\left\vert \boldsymbol{R}_{i,k^{\prime}}-\boldsymbol{R}_{i,k}\right\vert
=\frac{a}{K}|k-k^{\prime}|\qquad k\neq k^{\prime}\qquad k\ne k^{\prime
}=1,\ldots,K-1\qquad i=1,\ldots,N\label{consttwoa}%
\end{equation}
Let's consider now Eq.~(\ref{consttwo}) for
$k^{\prime}=K$, $k=1,\ldots,K-1$ and $i=1,\ldots,N-1$.  
This set of conditions contains
the coordinates $\boldsymbol R_{i,K}$ for $i=1,\ldots,N-1$. They
should be substituted by the
variables $\boldsymbol R_{i+1,1}$'s as
mentioned above. This operation results in the new conditions:
%Keeping the degrees of freedom  $\boldsymbol R_{i+1,1}$'s and
%eliminating the $\boldsymbol R_{i,K}$'s
%the conditions (\ref{consttwo}) become in this range of indexes:
%\begin{equation}
%\left\vert \boldsymbol{R}_{i,K}-\boldsymbol{R}_{i,k}\right\vert =\frac{a}%
%{K}(K-k)\qquad i=1,...,N-1\qquad k=1,...,K-1\label{fghlm}%
%\end{equation}
%We wish to eliminate the $\boldsymbol{R}_{i,k}$'$s$ using the conditions
%(\ref{constone}), so that Eq, (\ref{fghlm}) becames:%
\begin{equation}
\left\vert \boldsymbol{R}_{i+1,1}-\boldsymbol{R}_{i,k}\right\vert =\frac{a}%
{K}|K-k|\qquad i=1,\ldots,N-1\qquad k=1,\ldots,K-1\label{consttwob}%
\end{equation}
At this point, we are left only with the subset of
equations (\ref{consttwo}) which is strictly
related to the spheres of the $N-$th bar.
Clearly, the $K-$th sphere in the $N-$th bar is on one of the two free
ends of the chain. Thus, it is not constrained by
Eq. (\ref{constone}) and
the constraints of  Eq.~(\ref{consttwo}) should not be changed in this case:
\begin{equation}
\left\vert \boldsymbol{R}_{N,K}-\boldsymbol{R}_{N,k}\right\vert =\frac{a}%
{K}|K-k|\qquad k=1,\ldots,K-1\label{consttwoc}%
\end{equation}
%The advantage of having 
%expressed the constraints
%(\ref{consttwo}) in the form
% (\ref{consttwoa}),
%(\ref{consttwob}) and (\ref{consttwoc})
%is that in the new formulation the constraints (\ref{constone}) are
%automatically implemented and can be forgotten.

We are now ready to proceed with the construction of the generating
functional of the FJBC.
Exactly as we did for the FHC, we first introduce   the Langevin
equations that describe the fluctuations of the spheres. In the
present situation the
number of spheres is $KN$. Their positions are denoted by the
vectors $R_{i,k}$, where $i=1,\ldots,N$ and $k=1,\ldots,K$.
Accordingly, we have to introduce a set of $KN$ Langevin equations:
\begin{equation}
\boldsymbol{\dot{R}}_{i,k}(t)=\boldsymbol{\nu}_{i,k}{\normalsize (t)}\qquad
i=1,\ldots,N\qquad k=1,\ldots,K\label{langenaiwe}%
\end{equation}
In analogy with the case of the FHC, 
the probability distributions of the noises $\boldsymbol{\nu}_{i,k}$'s
are nonlinear
as a  consequence of the constraints of
Eqs.~(\ref{consttwoa}--\ref{consttwoc}). Before giving the explicit
expressions of these 
distributions, however, we have
to impose in equations (\ref{langenaiwe}) the constraints
(\ref{constone}). Due to these constraints, in fact,
not all the random forces appearing in
Eq.~(\ref{langenaiwe}) are independent.
% , which have been eliminated by reducing
%the number of degrees of freedom $R_{i,k}$. 
%As a consequence, we need also to reduce accordingly the number of
%noise sources.
Indeed, using the relations (\ref{constone}) it is easy to check that:
\begin{equation}
\boldsymbol{\nu}_{i,K}=\boldsymbol{\nu}_{i+1,1}\qquad
i=1,\ldots,N-1\label{noiseconstraints} 
\end{equation}
The above equations have a very simple physical explanation.
In order to join the bars together, some of the spheres have been
identified. 
As a result, the spheres identified in this way are also subjected to the
same noise and thus Eqs.~(\ref{noiseconstraints}) should be satisfied.
In a similar way as we did for the
constraints (\ref{consttwo}),
we may easily get rid of the redundant noises
$\boldsymbol{\nu}_{i,K}$, $i=1,\ldots,N-1$
 from the Langevin
equations (\ref{langenaiwe})  at the price of dividing these equations
into two sets. In the first set there will be only the spheres with
indexes $k<K$:
\begin{equation}
\boldsymbol{\dot{R}}_{i,k}=\boldsymbol{\nu}_{i,k}\qquad i=1,\ldots,N\qquad
k=1,\ldots,K-1\label{langeonea}%
\end{equation}
The degrees of freedom
 $\boldsymbol R_{i,K}$ and   $\boldsymbol{\nu}_{i,K}$ for
$i=1,\ldots,N-1$ 
are spurious due to the constraints  (\ref{constone})
and (\ref{noiseconstraints}), so that they should not be taken into account.
It remains only to consider the fluctuations of
the
last sphere
on the last bar corresponding to the indexes $i=N$ and $k=K$. This sphere
is not constrained
because it is located at one free end of the chain.
The related  Langevin equation reads as follows:
\begin{equation}
\boldsymbol{\dot{R}}_{N,K}=\boldsymbol{\nu}_{N,K}\label{langetwoa}%
\end{equation}
There are no other 
independent Langevin equations besides
Eqs. (\ref{langeonea}) and (\ref{langetwoa}). 

Having eliminated the constraints
(\ref{constone}), we are ready to construct the generating functional
of the FJBC. Let us note that the remaining constraints
are imposing conditions on the 
reciprocal distances between two points (the
centers of mass of the  spheres)
 exactly as the constraints of the FHC given in
Eq.~(\ref{constrequ}) do. 
As a consequence, it is possible to use the same strategy proposed in
Subsection~\ref{fhccase} in order to fix
the conditions  of 
Eqs.~(\ref{consttwoa}--\ref{consttwoc}).
Accordingly, we introduce the following non-gaussian
noise distribution:
\begin{eqnarray}
\mathcal{D}\rho(\{
\boldsymbol\nu_{i,k}\})&=&\left[\prod_{i=1}^N\prod_{k=1}^{K-1}
\mathcal{D}\boldsymbol{\nu}_{i,k}
\right]{\cal D}\boldsymbol{\nu}_{N,K}
\exp\left[  -\frac{1}{4D}
\sum_{i=1}^N\sum_{k=1}^{K-1}
\int_{0}^{t_{f}}\boldsymbol{\nu}_{i,k}^{2}dt\right]\nonumber\\
&\times&\exp\left[
-\frac{1}{4D}\int_{0}^{t_{f}}\boldsymbol{\nu}_{N,K}^{2}dt
\right]e^{-\tilde{W}(\{
\boldsymbol R_{i,k,\boldsymbol\nu_{i,k}}
\},\beta_1,\beta_2,\beta_3)}\label{noisedistfjbc}
%\qquad\qquad\left\{
%\begin{array}{ll}
%i=1,\ldots,N;&\qquad k=1,\ldots,K-1\\
%i=N;&\qquad k=K
%\end{array}
%\right.
\end{eqnarray}
The potential $\tilde{W}(\{
\boldsymbol R_{i,k,\boldsymbol\nu_{i,k}}
\},\beta_1,\beta_2,\beta_3)$ is a function of the solutions $\{
\boldsymbol R_{i,k,\boldsymbol\nu_{i,k}}
\}$ of Eqs.~(\ref{langeonea}--\ref{langetwoa}) and thus of the
noises $\boldsymbol\nu_{i,k}$ for $i=1,\ldots,N-1$ and
$k=1,\ldots,K-1$ and $\boldsymbol\nu_{N,K}$.
The form of $\tilde{W}$
can be chosen
following
the same criteria \ref{criter1}.~--~\ref{criter3}.
discussed in Subsection~\ref{fhccase}. 
A slight difference from the FHC is that, due to
the splitting of the constraints (\ref{consttwo}) into the three sets
of equations
given in (\ref{consttwoa}--\ref{consttwoc}), the FJBC potential
$\tilde{W}(\{
\boldsymbol R_{i,k,\boldsymbol\nu_{i,k}}
\},\beta_1,\beta_2,\beta_3)$ will depend on three parameters
$\beta_1,\beta_2$ and $\beta_3$
instead of one. The underlying idea is however the same.
When these parameters approach infinity,  $\tilde{W}(\{
\boldsymbol R_{i,k,\boldsymbol\nu_{i,k}}
\},\beta_1,\beta_2,\beta_3)$ should be chosen in such a way that it
becomes infinite outside  the 
region in the coordinate space in which the constraints
(\ref{consttwoa}--\ref{consttwoc}) are satisfied and zero otherwise.
Let's now derive  $\tilde{W}(\{
\boldsymbol R_{i,k,\boldsymbol\nu_{i,k}}
\},\beta_1,\beta_2,\beta_3)$ explicitly.
We have to require that the solutions $\boldsymbol
R_{i,k,\boldsymbol\nu_{i,k}}$ of the Langevin equations
(\ref{langeonea})
and (\ref{langetwoa}) satisfy
the constraints 
(\ref{consttwoa}--\ref{consttwoc}).
We recall that the
$\boldsymbol R_{i,k,\boldsymbol\nu_{i,k}}$'s
are purely functions of the noises $\nu_{i,k}$, so that the relations
(\ref{consttwoa}--\ref{consttwoc}) are conditions
on the noise degrees of freedom. 
Moreover,  Eqs.~(\ref{consttwoa}--\ref{consttwoc}) are of the form:
\begin{equation}
\boldsymbol R_a-\boldsymbol R_b=d_{ab}\label{constrondist}
\end{equation}
In other words, as previously stressed, they constrain the distance between
two points $a$ i $b$ exactly as the constraints of the FHC of
Eq.~(\ref{constrequnu}). 
In order to implement constraints of this kind we can imagine,
as it has been done in the previous Subsection,  that the two points
$a$ and $b$ are connected together by a spring with rest 
length $d_{ab}>0$. In the
limit of infinite elastic constant, the spring is frozen in its
rest position and the distance between the points is fixed to
$d_{ab}$.
These consideration suggest that constraints of the kind given in
Eq.~(\ref{constrondist}) may be fixed using the elastic potential
\begin{equation}
V_{ab}(|\boldsymbol R_a-\boldsymbol R_b|)=\frac\beta2\left(
|\boldsymbol R_a-\boldsymbol R_b|-d_{ab}
\right)^2\label{fhcpot}
\end{equation}
There is however no reason for restricting ourselves to elastic
interactions. Any two-body interaction that freezes the distance
between two points in the limit of infinite strength is suitable.
%
%Any potential that has an infinitely sharp peak in the region for
%which $|\boldsymbol R_a-\boldsymbol R_b|=d_{ab}$ is suitable.
Of course, the related potential 
should have  an infinitely sharp peak in the region for
which $|\boldsymbol R_a-\boldsymbol R_b|=d_{ab}$. Moreover, it
has to be
of the form $V(|\boldsymbol R_a-\boldsymbol
R_b|)=f^2(|\boldsymbol R_a-\boldsymbol R_b|)$, where $f$ is an
arbitrary function, otherwise it will no
longer be possible to use the identity (\ref{delide}).
For example, a valid potential is the following:
\begin{equation}
V'_{ab}|\boldsymbol R_a-\boldsymbol R_b|)=\beta\left(
|\boldsymbol R_a-\boldsymbol R_b|^2-d^2_{ab}
\right)^2\label{altpot}
\end{equation}
It is  easy to show that the above
potential is completely equivalent to the elastic 
one given in Eq.~(\ref{fhcpot}). This becomes clear if we rewrite
$V'_{ab}(|\boldsymbol R_a-\boldsymbol R_b|)$ in the form
\begin{equation}
V'_{ab}(|\boldsymbol R_a-\boldsymbol R_b|)=
\beta\left(
|\boldsymbol R_a-\boldsymbol R_b|-d_{ab}
\right)^2\left(
|\boldsymbol R_a-\boldsymbol R_b|+d_{ab}
\right)^2\label{ccc}
\end{equation}
Of course, the minimum of $V'_{ab}(|\boldsymbol R_a-\boldsymbol R_b|$
at $|\boldsymbol
R_a-\boldsymbol R_b|+d_{ab}=0$ 
will never be reached because $d_{ab}>0$.
Therefore, exactly as in the case of the elastic potential
(\ref{fhcpot}), $V'_{ab}(|\boldsymbol R_a-\boldsymbol R_b|)$ 
imposes only the condition  $|\boldsymbol R_a-\boldsymbol R_b|-d_{ab}=0$.
Moreover, it is possible to check that $V'_{ab}(|\boldsymbol
R_a-\boldsymbol R_b|)$ verifies all
conditions 1.~--~3. of Subsection~\ref{fhccase}. In
particular, in the 
limit $\beta=+\infty$ $V'_{ab}(|\boldsymbol R_a-\boldsymbol R_b|)$ is zero
if the constraints are satisfied and infinite in the opposite
case. The advantage of choosing the potential of Eq.~(\ref{altpot})
is that it leads directly to a delta
function in the desired form $\delta(|\boldsymbol R_a-\boldsymbol
R_b|^2-d_{ab}^2)$, so that we do not have to pass through all the
intermediate steps performed in Eqs.~(\ref{deltafunidenonfun}) and
(\ref{adfdhfj}) in order to arrive at the final generating functional
$Z_{FHC}[J]$ of the FHC given in Eq.~(\ref{genfunfjcdiscfinplus}).
Taking into account all the above considerations, to implement the
constraints of Eqs.~(\ref{consttwoa}--\ref{consttwoc}) we will choose
the following potentials:
%
%Having eliminated the coordinates and noises of the redundant beads,
%it is possible 
%to define 
%the noise distributions as follows:
%\begin{equation}
%\mathcal{D}\rho=\mathcal{D}\boldsymbol{\nu}_{i,k}\exp\left[  -\frac{1}{4D}%
%\int_{0}^{t_{f}}\boldsymbol{\nu}_{i,k}^{2}dt\right]
%\qquad\qquad\left\{
%\begin{array}{ll}
%i=1,\ldots,N;&\qquad k=1,\ldots,K-1\\
%i=N;&\qquad k=K
%\end{array}
%\right.
%\end{equation}
%To obtain the probability function of the FJBC we deform the
%probability function of a system of free beads by adding a suitable
%potential $\tilde{W}(\{R_{i,k}\})$ that is able to enforce the
%constraints (\ref{consttwoa}), (\ref{consttwob}) and
%(\ref{consttwoc}). These constraints look more complicated that those
%of the FHC, but they play the same role, in the sense that they are
%imposing conditions on the distances between the beads too.
%As a consequence, in analogy with what has been done in the case of he
%FHC,
%suitable potentials to fix the constraints will be elastic potentials
%whose minimum is obtained when the distances between the beads are
%those specified by Eqs.~(\ref{consttwoa}), (\ref{consttwob}) and
%(\ref{consttwoc}). The potentials corresponding to these constraints
%are respectively given by:
\begin{equation}
V_{i,kk^{\prime}}(\left\vert \boldsymbol{r}_{i,k}-\boldsymbol{r}_{i,k^{\prime
}}\right\vert )=(\Delta l)^2\beta_{1}\left(
\frac{\left\vert \boldsymbol{r}% 
_{i,k}-\boldsymbol{r}_{i,k^{\prime}}\right\vert ^{2}}{a^{2}}-\frac{\left\vert
k-k^{\prime}\right\vert ^{2}}{K^{2}}\right)  ^{2}\qquad 
\left\{\begin{array}{ccc}
i&=&1,\ldots,N-1\\
k\ne k^{\prime}&=&1,\ldots,K-1
\end{array}\right.\label{potentialone}
\end{equation}

\begin{equation}
V_{i,k}(\left\vert \boldsymbol{r}_{i+1,1}-\boldsymbol{r}_{i,k}\right\vert
)=\Delta l\beta_{2}\left(  \frac{\left\vert \boldsymbol{r}_{i+1,1}-\boldsymbol{r}%
_{i,k}\right\vert ^{2}}{a^{2}}-\frac{\left\vert K-k\right\vert ^{2}}{K^{2}%
}\right)  ^{2}\qquad
\left\{
\begin{array}{ccc}
 i&=&1,\ldots,N-1\\
 k&=&1,\ldots,K-1
\end{array}
\right.\label{potentialtwo}
\end{equation}

\begin{equation}
V_{k}(\left\vert \boldsymbol{r}_{N,K}-\boldsymbol{r}_{N,k}\right\vert
)=\Delta l\beta_{3}\left(  \frac{\left\vert \boldsymbol{r}_{N,K}-\boldsymbol{r}%
_{N,k}\right\vert ^{2}}{a^{2}}-\frac{\left\vert K-k\right\vert ^{2}}{K^{2}%
}\right)  ^{2}\qquad k=1,\ldots,K-1\label{potentialthree}
\end{equation}
Here $\beta_{1}$, $\beta_{2}$, $\beta_{3}$ are real parameters and are
supposed to be very large. Eventually, we will take the limit $\beta_{1}$,
$\beta_{2}$, $\beta_{3}\rightarrow\infty$. 
Let us note the change of length scale with respect to the FHC. Here
the smallest scale is the distance $\Delta l$ 
between the centers of mass of two contiguous spheres belonging to the
same bar.
As a consequence, the analog of Eq.~(\ref{potclasfhc}) in the present case is:
\begin{equation}
V_{FJBC}(\Delta l)=\sum_{k=1}^K\Delta l {\tilde V}_E(\boldsymbol R_{i,k}(t)) +
\sum_{\underset{k\ne k'}{k,k'=1}}^K\Delta l^2 {\tilde V}_I(\boldsymbol
R_{i,k}(t),\boldsymbol R_{j,k'}(t))\label{potdensfjbc}
\end{equation}
The above form of the potential guarantees the correct passage
to the limit $\Delta l\to 0$ and $K\to+\infty$ in which the bars
become continuous and one-dimensional systems with uniform mass
distribution. 
Indeed, it may be easily checked that in this continuous limit
Eq.~(\ref{potdensfjbc}) reduces to the potential for a chain of
one-dimensional bars with uniform and continuous mass distribution
given in
Eq.~(\ref{potdensfjbccont}). 
%Let us note how $a$ emerges as the new
%smallest scale from the 
%Eq.~(\ref{ascaleemerg}) after  
%of distances in a natural way from the continuous in Eq.~(\ref{ascaleemerg}).
%when one takes the continuous chain limit $a\to 0$ and $N\to+\infty$
%as it is shown by 

Going back to the main problem of constructing the generating
functional of the FJBC, we define $\tilde{W}(\{R_{i,k}\})$ 
as the  following linear combination of the
potentials (\ref{potentialone}--\ref{potentialthree}):
\begin{eqnarray}
\tilde{W}(\{r_{i,k}\})&=&\sum_{i=1}^{N}\sum_{k,k^{\prime}=1\atop k\ne
  k'}^{K-1}\int_{0}^{t_{f}}\left[ 
V_{i,kk^{\prime}}(\left\vert \boldsymbol{r}_{i,k}-\boldsymbol{r}_{i,k^{\prime
}}\right\vert )+\sum_{i=1}^{N-1}\sum_{k=1}^{K-1}V_{i,k}(\left\vert
\boldsymbol{r}_{i+1,1}-\boldsymbol{r}_{i,k}\right\vert )
\right.\nonumber\\
&+&\left.\sum_{k=1}^{K-1}%
V_{k}(\left\vert \boldsymbol{r}_{N,K}-\boldsymbol{r}_{N,k}\right\vert
)\right]  dt\label{potwtil}
\end{eqnarray}
%appearing in the exponential at the end of Eq.~(\ref{zfjbcconstr}).

By substituting the right hand side of Eq.~(\ref{potwtil})
 in Eq.~(\ref{noisedistfjbc}), we obtain
the noise distribution $\mathcal{D}\rho(\{
\boldsymbol\nu_{i,k}\})$. Knowing the noise distribution it is
possible
to construct the generating functional of the 
FJBC in path integral form:
\begin{align}
Z_{FJBC}[J]  & =\lim_{\beta_1,\beta_2,\beta_3\to+\infty}\left[
  \prod_{i=1}^{N}\prod_{k=1}^{K-1}\int\mathcal{D}% 
\boldsymbol{\nu}_{i,k}\right]  \exp\left[  -
\sum_{i=1}^{N}
\sum_{k=1}^{K-1}\int_{0}^{t_{f}}\left(
\frac{1}{4D}\boldsymbol{\nu}_{i,k}^{2}+\Delta
l\boldsymbol{J}_{i,k}\cdot\boldsymbol{R}_{i,k,\boldsymbol{\nu}_{i,k}}\right)dt\right] 
\nonumber\\
&  \int\mathcal{D}\boldsymbol{\nu}_{N,K} \exp\left[  -
\int_{0}^{t_{f}}\left(
\frac
{1}{4D}
\boldsymbol{\nu}_{N,K}^{2}
+\Delta
l\boldsymbol{J}_{N,K}\cdot\boldsymbol{R}_{N,K,\boldsymbol{\nu}_{N,K}}\right)dt\right]
\exp\left[- {\tilde W}(\{R_{i,k,\boldsymbol \nu_{i,k}}\})\right]  \label{zfjbcconstr} 
\end{align}

At this point, in analogy with the steps (\ref{step1}--\ref{step2})
made in Section~\ref{fhccase},
 we insert in the generating functional
(\ref{zfjbcconstr}) the quantity:
\begin{equation}
\tilde{I}=\left[  \prod_{i=1}^{N}\prod_{k=1}^{K-1}\int\mathcal{D}%
\boldsymbol{r}_{i,k}\delta(\boldsymbol{r}_{i,k}-\boldsymbol{R}%
_{i,k,\boldsymbol{\nu}_{i,k}})\right]  \int\mathcal{D}\boldsymbol{r}%
_{N,K}\delta(\boldsymbol{r}_{N,K}-\boldsymbol{R}_{N,K,\boldsymbol{\nu}_{N,K}})
\end{equation}
Clearly $\tilde{I}=1$, so its insertion 
in Eq.~(\ref{zfjbcconstr})
does not change the physics of the
problem. It is also possible to check that:
\begin{equation}
\tilde{I}=\left[  \prod_{i=1}^{N}\prod_{k=1}^{K-1}\det\left[  \frac{\partial
}{\partial t}\right]  \int\mathcal{D}\boldsymbol{r}_{i,k}\delta
(\boldsymbol{\dot{r}}_{i,k}-\boldsymbol{\nu}_{i,k})\right]  \det\left[
\frac{\partial}{\partial t}\right]  \int\mathcal{D}\boldsymbol{r}_{N,K}%
\delta(\boldsymbol{\dot{r}}_{N,K}-\boldsymbol{\nu}_{N,K})
\end{equation}
As a result of the insertion of $\tilde{I}$, we get after an
easy integration over the noises $\boldsymbol{\nu}_{i,k}$ and
$\boldsymbol{\nu}_{N,K}$:%
\begin{eqnarray}
Z_{FJBC}[J]  & =& \lim_{\beta_1,\beta_2,\beta_3\to+\infty}\left[
  \prod_{i=1}^{N}\prod_{k=1}^{K-1}\int\mathcal{D} 
\boldsymbol{r}_{i,k}\right]  \exp\left[  -\sum_{i=1}^N\sum_{k,k^{\prime}=1\atop k\ne
  k'}^{K-1} 
\int_{0}^{t_{f}}\left(
\frac{1}{4D}
\boldsymbol{\dot{r}}_{i,k}^{2}+
\Delta l\boldsymbol{J}_{i,k}\cdot\boldsymbol{r}_{i,k}\right)dt\right]
\nonumber\\ 
&\times&   \int\mathcal{D}\boldsymbol{r}_{N,K}
\exp\left[  -
\int_{0}^{t_{f}}\left(\frac 
{1}{4D}
\boldsymbol{\dot{r}}_{N,K}^{2}dt
+\Delta l\boldsymbol{J}_{N,K}\cdot\boldsymbol{r}_{N,K}\right)dt\right]
\nonumber\\ 
&\times& \exp\left[
  -(\Delta l)^2\beta_{1}\sum_{i=1}^{N}\sum_{k,k^{\prime}=1\atop k\ne
    k'}^{K-1}\int 
_{0}^{t_{f}}\left(  \frac{\left\vert \boldsymbol{r}_{i,k}-\boldsymbol{r}%
_{i,k^{\prime}}\right\vert ^{2}}{a^{2}}-\frac{\left\vert k-k^{\prime
}\right\vert ^{2}}{K^{2}}\right)  ^{2}dt\right]  \nonumber\\
&\times& \exp\left[
  -\Delta l\beta_{2}\sum_{i=1}^{N-1}\sum_{k=1}^{K-1}\int_{0}^{t_{f}% 
}\left(  \frac{\left\vert \boldsymbol{r}_{i+1,1}-\boldsymbol{r}_{i,k}%
\right\vert ^{2}}{a^{2}}-\frac{\left\vert K-k\right\vert ^{2}}{K^{2}}\right)
^{2}dt\right]  \nonumber\\
&\times &\exp\left[  -\Delta l\beta_{3}\sum_{k=1}^{K-1}\int_{0}^{t_{f}}\left(
\frac{\left\vert \boldsymbol{r}_{N,K}-\boldsymbol{r}_{N,k}\right\vert ^{2}%
}{a^{2}}-\frac{\left\vert K-k\right\vert ^{2}}{K^{2}}\right)  ^{2}dt\right]
\label{zfjbclmn}
\end{eqnarray}

Finally, we apply Eq. (\ref{delide}) in order to perform the limits
$\beta_{1}$, $\beta_{2}$, $\beta_{3}\rightarrow+\infty$. The result is:%
\begin{eqnarray}
Z_{FJBC}[J]  & =&
\left[
  \prod_{i=1}^{N}\prod_{k=1}^{K-1}\int\mathcal{D} 
\boldsymbol{r}_{i,k}\right]  \exp\left[  -\sum_{i=1}^N\sum_{k,k^{\prime}=1\atop k\ne
  k'}^{K-1} 
\int_{0}^{t_{f}}\left(
\frac{1}{4D}
\boldsymbol{\dot{r}}_{i,k}^{2}+
\Delta l\boldsymbol{J}_{i,k}\cdot\boldsymbol{r}_{i,k}\right)dt\right]
%
%\left[  \prod_{i=1}^{N}\prod_{k=1}^{K-1}\int\mathcal{D}
%\boldsymbol{r}_{i,k}\right]  \exp\left[  -\frac{1}{4D}\int_{0}^{t_{f}
%}\boldsymbol{\dot{r}}_{i,k}^{2}dt\right]  \exp\left[  -\int_{0}^{t_{f}
%}\boldsymbol{J}_{i,k}\cdot\boldsymbol{r}_{i,k}dt\right]
\nonumber  \\
&\times&   \int\mathcal{D}\boldsymbol{r}_{N,K}
\exp\left[  -
\int_{0}^{t_{f}}\left(\frac 
{1}{4D}
\boldsymbol{\dot{r}}_{N,K}^{2}dt
+\Delta l\boldsymbol{J}_{N,K}\cdot\boldsymbol{r}_{N,K}\right)dt\right]
%&\times& \left[  \int\mathcal{D}\boldsymbol{r}_{N,K}\right]
%\exp\left[  -\frac 
%{1}{4D}\int_{0}^{t_{f}}\boldsymbol{\dot{r}}_{N,K}^{2}dt\right]  \exp\left[
%-\int_{0}^{t_{f}}\boldsymbol{J}_{N,K}\cdot\boldsymbol{r}_{N,K,}dt\right]
\nonumber\\
&\times& \left[  \prod\limits_{i=1}^{N}\prod\limits_{\substack{k,k^{\prime
}=1\\k\neq k^{\prime}}}^{K-1}\delta\left(  \frac{\left\vert \boldsymbol{r}%
_{i,k}-\boldsymbol{r}_{i,k^{\prime}}\right\vert ^{2}}{a^{2}}-\frac{\left\vert
k-k^{\prime}\right\vert ^{2}}{K^{2}}\right) \right]  \nonumber\\
&\times& \left[
  \prod\limits_{i=1}^{N-1}\prod\limits_{k=1}^{K-1}\delta\left( 
\frac{\left\vert \boldsymbol{r}_{i+1,1}-\boldsymbol{r}_{i,k}\right\vert ^{2}
}{a^{2}}-\frac{\left\vert K-k\right\vert ^{2}}{K^{2}}\right)  \right]
\nonumber\\
&\times& \left[  \prod\limits_{k=1}^{K-1}\delta\left(
  \frac{\left\vert 
\boldsymbol{r}_{N,K}-\boldsymbol{r}_{N,k}\right\vert ^{2}}{a^{2}}%
-\frac{\left\vert K-k\right\vert ^{2}}{K^{2}}\right)
\right] \label{gengenfundisc} 
\end{eqnarray}
This is the expression of the generating functional of the
discrete FJBC in the shish kebab approximation.
%Before passing to  the continuous limit, it is necessary to provide a
%precise 
%recipe in order to deal with the two scales of lengths $\Delta l$ and
%%$a$. 
%We should decide which one will be taken as the fundamental scale of
%length. 
%We note that when the entire length $a$ of each bar goes to zero, it
%makes 
%no 
%sense to divide the bars into $K$ different elements. For this reason,
%in the 
%limit $a\rightarrow0$ we will keep $a$ as the fundamental length
%scale and 
%divide the bar into a minimal set of elements. 
\section{The double continuous limit of the FJBC}\label{section:four}
Before performing the continuous limit of the generating functional
(\ref{gengenfundisc}), it is important to recall how the FJBC is
constructed.
The
FJBC consists in one-dimensional systems with an uniform and
continuous
distribution
of mass as it was explained when discussing the classical case and in
Subsection~\ref{subsection:FJBC}.
In our approach each bar has been approximated by a shish-kebab model,
in which $K$ small spheres are put together in a row by imposing the
constraints  (\ref{consttwoa}--\ref{consttwoc}). The distance between
the centers of mass of the spheres is given by the quantity $\Delta l$ of
Eq.~(\ref{distbetwspher}). 
One can see the spheres as small masses $m$ localized at regular
intervals on a segment of length $a$.
The three dimensionality of the spheres is adjusted
by tuning the diffusion constant $D$. In the present case $D$ has been
choosen in such a way that it coincides with the diffusion constant of
a small sphere of 
diameter $\Delta l$. Of course, if the length $a$ of the bar is finite,
the more spheres we will use, the better will be the approximation of the
bar.
To reproduce the FJBC, an infinite number of spheres
is necessary. Indeed, when $K$ is infinite, the diameter of the
spheres approaches zero as it is shown by Eq.~(\ref{distbetwspher})
and the shish-kebab models coincides with an uniform distribution of
point-like masses distributed on a segment of length $a$. This is
exactly the one-dimensional bar that has been taken as the basic unit
in the FJBC model. 
%In the continuous limit we can still let the number
%$K$ of spheres to be large, but actually this is not strictly
%necessary, because the bar reduces in this case to infinitesimal
%elements of continuous chain and its description by an arbitrary
%number of sphere becomes redundant if $K>2$.
%This fact may be checked directly. 

As a warming up exercise, the probability function of the FJBC will be
computed in the limit $\Delta l\to 0$ and $K\to+\infty$. As explained
before, this limit corresponds to the one-dimensional chain with
uniform mass distribution without the shish-kebab approximation.
In that limit $\tilde L$ and $L$ coincide and the length of each bar
will be equal to $a$ because the decrease of the  length due to the
way in which the bars are joined together becomes negligible when
$\Delta l\to 0$ and $K\to+\infty$.

First of all, we derive the expression of the mass $m$ of
a single sphere in terms of two macroscopic parameters, namely the total
length of the chain $\tilde L$ and its total mass $M$.
To compute $M$, we subtract from the total mass $NKm$ of $N$ bars the mass
$(N-1)m$ of the $N-1$ spheres which disappear because they are
identified
 at the junctions
between the bars. The result is $M=NKm-(N-1)m$,
Thus we may write:
\begin{equation}
m=\frac M{N(K-1)+1}
\end{equation}
Using Eq.~(\ref{totlengthdischain}), we get:
\begin{equation}
m=\frac{Ma}{K\tilde L}=\frac {M}{\tilde L}\Delta l
\end{equation}
where $\frac M{\tilde L}$ is the uniform mass
density. 
The analog of Eq.~(\ref{14D}) in the FJBC case is:
\begin{equation}
\frac 1{4D}=\frac{1}{2kT\tau}\frac
      {M}{2{\tilde L}}\Delta l\label{14Dfjbc} 
\end{equation}
Next, we provide the prescriptions to pass to the continuous bar limit
$\Delta l\to 0$ and $K\to+\infty$:
\begin{eqnarray}
\sum_{k=1}^{K-1}\Delta l\to \int_0^a
dl&\qquad\qquad&\sum_{k,k'=1}^{K-1}(\Delta
l)^2\to\int_0^adldl'\label{prescr1}\\ 
\boldsymbol r_{i,k}(t)\to\boldsymbol r_i(t,l)&\qquad\qquad&{\tilde
  L}\to L\label{prescr2}\\ 
\boldsymbol r_i(t,0)=\boldsymbol r_{i,1}(t)
&\qquad\qquad&\boldsymbol r_i(t,a)=\boldsymbol r_{i+1,1}(t)
\label{prescr3}\\
a\frac kK\to l&\qquad\qquad&\label{prescr4}
\end{eqnarray}
Using the above formulas (\ref{14Dfjbc}--\ref{prescr4}) is is not
difficult to show that:
\begin{eqnarray}
Z_{FJBC}^{K=\infty}[J]&=&\lim_{\beta_1,\beta_2,\beta_3\to+\infty}
\left[
\prod_{i=1}^N\int{\cal D}\boldsymbol r_i(t,l)
\right]\exp\left[
-\sum_{i=1}^N\int_0^{t_f}dtI_i^{(0)}(t)
\right]\nonumber\\
&\times&\exp\left[
-\beta_1\sum_{i=1}^N\int_0^{t_f}dt I_i^{(1)}(t)
\right]\exp\left[
-\beta_2\sum_{i=1}^{N-1}\int_0^{t_f}dt I_i^{(2)}(t)
\right]\nonumber\\
&\times&\exp\left[
-\beta_3\int_0^{t_f}dt I_i^{(3)}(t)
\right]\label{probfunklim}
\end{eqnarray}
where
\begin{equation}
I_i^{(0)}(t)=\int_0^adl\left(
\frac{1}{2kT\tau}\frac {M}{2L}\dot{\boldsymbol r}_i^2(t,l)+\boldsymbol
J_i(t,l) \cdot\boldsymbol r_i(t,l)
\right)\label{ii0}
\end{equation}
\begin{equation}
I_i^{(1)}(t)=\int_0^adldl'\left(
\frac{|\boldsymbol r_i(t,l)-\boldsymbol r_i(t,l')
  |^2}{a^2}-\frac{|l-l'|^2}{a^2} 
\right)^2\label{i1t}
\end{equation}
\begin{equation}
I_i^{(2)}(t)=\int_0^adl\left(
\frac{|\boldsymbol r_{i+1}(t,0)-\boldsymbol r_i(t,l)
  |^2}{a^2}-\frac{|a-l|^2}{a^2} 
\right)^2
\end{equation}
\begin{equation}
I_N^{(3)}(t)=\int_0^adl\left(
\frac{|\boldsymbol r_N(t,a)-\boldsymbol r_N(t,l)
  |^2}{a^2}-\frac{|a-l|^2}{a^2} 
\right)^2\label{ii3}
\end{equation}
By taking the limit $\beta_1,\beta_2,\beta_3\to+\infty$ we finally
obtain  the probability function of the FJBC: 
\begin{eqnarray}
Z_{FJBC}^{K=\infty}[J]&=&
\left[
\prod_{i=1}^N\int{\cal D}\boldsymbol r_i(t,l)
\right]
\exp\left[
-\sum_{i=1}^N\int_0^{t_f}dt\int_0^adl\left(
\frac{1}{2kT\tau}\frac {M}{2L}\dot{\boldsymbol r}_i^2(t,l)+\boldsymbol
J_i(t,l) \cdot\boldsymbol r_i(t,l)
\right)
\right]\nonumber\\
&\times&\left[
\prod_{i=1}^N\delta\left(
\frac{|\boldsymbol r_i(t,l)-\boldsymbol r_i(t,l')
  |^2}{a^2}-\frac{|l-l'|^2}{a^2} 
\right)
\right]
\left[
\prod_{i=1}^{N-1}\delta\left(
\frac{|\boldsymbol r_{i+1,1}(t)-\boldsymbol r_i(t,l)
  |^2}{a^2}-\frac{|a-l|^2}{a^2} 
\right)
\right]\nonumber\\
&\times&
\delta\left(
\frac{|\boldsymbol r_N(t,a)-\boldsymbol r_N(t,l)
  |^2}{a^2}-\frac{|a-l|^2}{a^2} 
\right)\label{probfunklimfjbc}
\end{eqnarray}
We would like to check if, in the limit of a continuous chain
(\ref{contlim}), the generating functional of the FJBC reduces to the
partition function of the GNL$\sigma$M of Eq.~(\ref{genfunfjc}) as it
should be expected. Our starting point will be the generating
functional $Z_{FJBC}^{K=\infty}[J]$ in the version of
Eq.~(\ref{probfunklim}). First of all we expand the functions
$I_i^{(0)}(t),\ldots,I_N^{(3)}(t)$  given in
Eqs.~(\ref{ii0}--\ref{ii3}) at
the leading order in $a$. Since at the end the limit $a\to0$ will be
taken, there is no need to compute higher order terms.
A straightforward application of the expansion in Taylor series gives
for $I_i^{(0)}(t)$:
\begin{equation}
I_i^{(0)}(t)=\left(
\frac 1{2kT\tau}\frac M {2L}\dot{\boldsymbol r}^2_{i,1}(t)+\boldsymbol
J_{i,1}(t) \cdot\boldsymbol r_{i,1}(t)
\right)a\label{simpzero}
\end{equation}
To write down the above equation we have used the prescription
(\ref{prescr3}) according to which $\boldsymbol r_i(t,0)=\boldsymbol
r_{i,1}(t)$. Let us consider now the term $I_i^{(1)}(t)x$.
Due to the fact that we are working in the region of small values of
$a$, it is possible to make the 
following approximation:
\begin{equation}
\left|
\frac{\boldsymbol r_i(t,l)-\boldsymbol r_i(t,l')}{l-l'}
\right|\sim \left|
\frac{\boldsymbol r_i(t,a)-\boldsymbol r_i(t,a)}{a}
\right|\label{derivsimplif}
\end{equation}
Both members of the above equation coincide in fact with the modulus
of the derivative $\left|
\frac{\partial \boldsymbol r_i(t,0)}{\partial l}
\right|$ up to higher order terms in the infinitesimal quantities
$l,l'$ and $a$. Using Eq.~(\ref{derivsimplif}) to simplify the
expression of $I_i^{(1)}(t)$ given in (\ref{i1t}), we obtain:
\begin{equation}
I_i^{(1)}(t)\sim\left\{
\begin{array}{cc}
\left(
\frac{|\boldsymbol r_{i+1,1}(t)-\boldsymbol r_{i,1}(t)|^2}{a^2}
-1\right)^2A&\mbox{for $i=1,\ldots,N-1$}\\
\left(
\frac{|\boldsymbol r_{N}(t,a)-\boldsymbol r_{N}(t,0)|^2}{a^2}
-1\right)^2A&\mbox{for $i=N$}
\end{array}
\right.
\end{equation}
where $A=\int_0^adldl'\frac{|l-l'|^4}{a^4}$. A direct calculation
shows that:
\begin{equation}
A=\frac {a^2}{15}
\end{equation}
Similar calculations allow the computation of $I_i^{(2)}(t)$ and
$I_N^{(3)}(t)$:
\begin{equation}
I_i^{(2)}(t)\sim\left(
\frac{|\boldsymbol r_{i+1,1}(t)-\boldsymbol r_{i,1}(t)|^2}{a^2}
-1\right)^2\frac a5
\qquad\qquad i=1,\ldots,N-1
\label{simptwo}
\end{equation}
\begin{equation}
I_N^{(3)}(t)\sim\left(
\frac{|\boldsymbol r_{N}(t,a)-\boldsymbol r_{N}(t,0)|^2}{a^2}
-1\right)^2\frac a5\label{simpthree}
\end{equation}
As we see from Eqs.~(\ref{simpzero}--\ref{simpthree}), in the action of
the generating functional $Z_{FJBC}^{K=\infty}[J]$ many degrees of
freedom disappear when $a$ becomes small. The only remaining degrees
of freedom are the variables $\boldsymbol r_{i,1}(t)=\boldsymbol r_i(t,0)$ for
$i=1,\ldots,N$ and $\boldsymbol r_N(t,a)$. All
the other degrees of freedom $\boldsymbol r_i(t,l)$ for $l\ne 0$ and
$l\ne a$ can be simply integrated out in Eq.~(\ref{probfunklim}),
because they do not appear in the functions
$I_i^{(0)}(t),\ldots,I_N^{(3)}(t)$ of Eqs.~(\ref{ii0}--\ref{ii3}) and
thus in the
generating functional  $Z_{FJBC}^{K=\infty}[J]$. 
Their integration
will result in a constant overall factor $\cal C$ multiplied with the
rest of the generating functional.
Moreover, when $a$ goes to zero, all terms 
containing positive powers of 
$a$ vanish identically if these powers are not absorbed by an
appropriate number of sums over all bars according to the prescription
$\sum_{i=1}^Na\to \int_0^Lds$. In particular it is easy to check that
\begin{equation}
\lim_{a\to 0}\sum_{i=1}^N\int_0^{t_f}dt I_i^{(1)}=0
\end{equation}
and
\begin{equation}
\lim_{a\to 0}\int_0^{t_f}dt I_N^{(3)}=0
\end{equation}
As a consequence, for small values of  $a$  the leading order contribution
to the generating functional
$Z_{FJBC}^{K=\infty}[J]$ is provided by:
\begin{eqnarray}
Z_{FJBC}^{K=\infty}[J]&\sim&
\lim_{\beta_2\to+\infty}{\cal C}\left[
\prod_{i=1}^N\int{\cal D}\boldsymbol r_{i,1}(t)
\right]\exp\left[
-\sum_{i=1}^Na\int\limits_0^{t_f}dt\left(
\textstyle \frac 1{2kT\tau}\frac M{2L}\dot{\boldsymbol r}^2_{i,1}(t)+\boldsymbol
J_{i,1}(t)\cdot \boldsymbol r_{i,1}(t)
\right)
\right]\nonumber
\\
&\times&\exp\left[
-\frac{\beta_2}{5}\sum_{i=1}^{N-1}a\int_0^{t_f}dt\left(
\frac{|\boldsymbol r_{i+1,1}(t)-\boldsymbol r_{i,1}(t)|^2}{a^2}-1
\right)^2
\right]
\end{eqnarray}
Now it is possible to proceed with the continuous limit
(\ref{contlim}) following the prescription of
Subsection~\ref{subsection:fhc}. The result is:
\begin{equation}
\lim_{a\to 0,N\to+\infty,Na=L}Z_{FJBC}^{K=\infty}[J]=Z'[J]
\end{equation}
where
\begin{eqnarray}
Z'[J]&=&\lim_{\beta_2\to+\infty}{\cal C}\int{\cal D}\boldsymbol
r(t,s)\exp
\left[
-\int_0^{t_f}dt\int_0^Lds\left(
\textstyle \frac 1{2kT\tau}\frac M{2L}\dot{\boldsymbol
  r}^2(t,s)+\boldsymbol J(t,s)\cdot\boldsymbol r(t,s)
\right)
\right]\nonumber\\
&\times&\exp\left[
-\frac{\beta_2}{5}\int_0^{t_f}dt\int_0^Lds\left(
\left|
\frac{\partial\boldsymbol r(t,s)}{\partial s}
\right|^2-1
\right)^2
\right]
\end{eqnarray}
Performing also the limit for infinitely large values of $\beta_2$, we
obtain up to an irrelevant proportionality constant:
\begin{eqnarray}
Z'[J]&=&\int{\cal D}\boldsymbol
r(t,s)\exp
\left[
-\int_0^{t_f}dt\int_0^Lds\left(
\textstyle \frac 1{2kT\tau}\frac M{2L}\dot{\boldsymbol
  r}^2(t,s)+\boldsymbol J(t,s)\cdot\boldsymbol r(t,s)
\right)
\right]\nonumber\\
&\times&\delta\left(\left|
\frac{\partial\boldsymbol r(t,s)}{\partial s}
\right|^2-1\right)\label{genfuncttildefinal}
\end{eqnarray}
This is exactly the partition function of the GNL$\sigma$M 
of Eq.~(\ref{genfunfjc}) as expected.

\section{Conclusions}\label{section:conclusions}
In this work an approach to the dynamics of an inextensible chain has
been proposed.
At the basis of that approach there is the observation that in a
constrained mechanical system fluctuating in some viscous
medium at constant
temperature, the random noise acting on the system is subjected to
constraints too.
Following this simple observation, the constraints have been imposed
here directly on the noise degrees of freedom. In this way the
complications of having to deal with an extended set of degrees of
freedom (the Lagrange multipliers) or with generalized coordinates,
are absent.
In the generating functional
 of the FHC
introduced in
Eq.~(\ref{zfjcdef}), 
the constraints have been fixed by means of potentials.
We recall at this point that even in the case in which
the strength of the potentials imposing the inextensibility
constraints are very large, numerical simulations show that the
fast modes related to the changes of the lengths of the basic
units still dominate over the slow modes
associated to the conformational changes of the chain
\cite{liverpool,peters}. 
To eliminate this problem, in the generating functional of the FHC
the limit in which the strengths of the
potential (\ref{pottot}) %and (\ref{potwtil}) 
becomes infinite has been
performed. The final expression of the generating functional of the
FHC in this limit is shown in Eq.~(\ref{genfunfjcdiscfinplus}).
% The
%analogous expression for the generating functional of the FJBC is
%displayed in Eq.~(\refl{probfunklimfjbc}).

In this work it has also been explored the strategy of
Ref.~\cite{liverpool}, in which it has been proposed that an
inextensible chain may be realized gluing together a set of
inextensible basic units like for instance one-dimensional bars with
uniform mass distribution. 
In Subsection~\ref{subsection:FJBC} this strategy has been explored
for a chain in which the bars have been replaced by the shish-kebabs
displayed in Fig.~\ref{figone}. Each shish-kebab
consists
of $K$ spheres that are aligned together in a row by means of suitable
constraints in order to form the
approximate shape of a bar.
The final expression of the 
generating functional $Z_{FJBC}[J]$ for this type of chain may be
found in
Eq.~(\ref{gengenfundisc}).
In the limit in which the distance $\Delta l$ between the centers of
mass of the 
spheres 
 vanishes identically and the number $K$ of spheres
becomes infinitely large while the product $K\Delta l $ remains finite,
the shish-kebabs become one-dimensional bars and one obtains the
model of a chain which has been called here FJBC.
Its generating functional in path integral form has been given in
Eq.~(\ref{probfunklimfjbc}).
As a check of the consistency of our approach it has
been verified that in the limit of a continuous chain both generating
functionals of the FHC and FJBC coincide. Indeed, 
after performing the continuous limit, one obtains respectively
the generating functionals of Eqs.~(\ref{genfunfjc}) and
(\ref{genfuncttildefinal}) that describe the same theory, namely the
GNL$\sigma$M of Ref.~\cite{FePaVi}.
The consistency of the approach presented in this work in order to tackle
the problem of the dynamics of an inextensible chain is also confirmed
by the results of Ref.~\cite{FePaVi4}. In has been shown there that
the generating functional of the FHC given in
Eq.~(\ref{genfunfjcdiscfinplus}) consists of a statistical sum over
all particle trajectories that satisfy the Langevin equations
(\ref{langefree}) together with the inextensibility constraints
(\ref{constrequ}) as it is expected.
In \cite{FePaVi2} it
has been verified that in the continuous case the statistical sum in
the GNL$\sigma$M contains
only the chain conformations for which $\dot{\boldsymbol
  R}(t,s)=\boldsymbol\nu(t,s)$ and $\left|
\partial\boldsymbol R(t,s)/\partial s
\right|=1$.  These equations are respectively the continuous
counterparts of Eqs.~(\ref{langefree}) and (\ref{constrequ}).
Additionally, we have explored in this work the Fokker-Planck
formulation of the dynamics of the FHC in
Eqs.~(\ref{schroed-like})--(\ref{forces}).

Finally, let us mention some issues that are still open. First of all,
the  hydrodynamic and self-avoiding forces are missing
in our approach.
To describe the dynamics of realistic chains
these interactions should be taken into account.
For this reason, up to now the 
 applications of the GNL$\sigma$M are limited to situations in
which the speed of the elements of the chain is low, so that the
hydrodynamics interactions can be neglected.
Let us note that the addition of nonlinear self-interactions acting on
the beads of a set of coupled oscillators like that described by
Eq.~(\ref{zfjcabc}) leads to interesting phenomena as described in
Refs.~\cite{Saridakis1,Saridakis2,Saridakis3}, where such kind of systems has
been studied.
Another important issue is to find suitable approximations that may
be used to evaluate field path integrals 
in the presence of functional Dirac
delta functions and nontrivial boundary conditions.
This is exactly the case of the generating functional of
Eq.~(\ref{genfunfjc}). So far, the partition function and the
so-called dynamic structure factor have been derived in the semiclassical
approximation \cite{FePaVi,FePaVi2}. One could simplify the theory even
further if it would be possible to extend
the well known 
approximation of the Dirac delta function with a gaussian function
also to
the functional Dirac delta function imposing the
inextensible constraints in the GNL$\sigma$M.
At the moment, this extension has been achieved only in the case
of the statistical mechanics of the FHC \cite{EdwGoo}.
The generalization of this result to dynamics is still work in progress.

\section{Acknowledgments}
{One of the authors, F.~F., is indebted to H. Arod\'z,
  Z. Jask\'olski, 
J. Paturej, M. Pi\c{a}tek, V. G. Rostiashvili and T. A. Vilgis for fruitful
discussions.
This research was supported in part by the National Science Foundation
under Grant No. NSF PHY05-51164.
}

\appendix
\section*{Appendix A -- Proof of Eq. (\ref{delide})}
We start from the quantity
\begin{equation}
A(\beta)=C(\beta)\exp\left[-a\beta\int_0^{t_f}
dtf^2(\boldsymbol r_i(t),\boldsymbol r_j(t))
\right]\qquad\qquad i,j=1,\ldots,N\label{defabeta}
\end{equation}
where $C(\beta)$ is a normalization constant.
We discretize the interval of time $[0,t_f]$ into $M$ smaller
intervals of length $b$. Of course, $Mb=t_f$.
$A(\beta)$ is recovered in the continuous limit $b\longrightarrow0$
and $M\longrightarrow+\infty$:
\begin{equation}
A(\beta)=\lim_{\underset{M\to +\infty}{b\to 0}} C_M(\beta)
\exp\left[-a\beta\sum_{m=1}^Mbf^2(\boldsymbol r_i(t_m),\boldsymbol
  r_j(t_m)\right] 
\end{equation}
Here $t_m=mb$ is the discretized time variable and
$C(\beta)=\lim_{\underset{M\to +\infty}{b\to 0}}C_M(\beta)$.
Let us put
\begin{equation}
\frac 1\epsilon=a\beta b
\end{equation}
and 
\begin{equation}
C_M(\beta)=\prod_{m=1}^M\frac 1{(4\pi\epsilon)^{\frac 12}}
\end{equation}
Thus
\begin{equation}
A(\beta)=\lim_{\underset{M\to +\infty}{b\to 0}}\left[\prod_{m=1}^M
\frac 1{(4\pi\epsilon)^{\frac 12}}
\exp{\left(-\frac { f^2(\boldsymbol r_i(t_m),\boldsymbol
    r_j(t_m))}{\epsilon}\right)}\right] 
\end{equation}
We consider at this point the limit $\beta\longrightarrow +\infty$ of
$A(\beta)$:
\begin{equation}
\lim_{\beta\to+\infty}A(\beta)=\lim_{\epsilon\to
  0}\lim_{\underset{M\to +\infty}{b\to 0}} \left[\prod_{m=1}^M
\frac 1{(4\pi\epsilon)^{\frac 12}}
\exp{\left(-\frac{ f^2(\boldsymbol r_i(t_m),\boldsymbol
    r_j(t_m))}{\epsilon}\right)} \right]
\end{equation}
Permuting the limit $\epsilon\to 0$ with the limit $M\to+\infty$ and
$b\to 0$, we get:
\begin{equation}
\lim_{\beta\to+\infty}A(\beta)=\lim_{\underset{M\to +\infty}{b\to 0}} 
\left[\prod_{m=1}^M\lim_{\epsilon\to
  0} 
\frac 1{(4\pi\epsilon)^{\frac 12}}
\exp{\left(-\frac{ f^2(\boldsymbol r_i(t_m),\boldsymbol
    r_j(t_m))}{\epsilon}\right)} \right]
\end{equation}
When $\beta$ starts to be very large and simultaneously $\epsilon$
becomes very small, it is possible to use the Gaussian
representation of the Dirac delta function
$\delta(x)=\lim_{\epsilon\to0}
\frac1{(4\pi\epsilon)^{\frac 12}}e^{-\frac{x^2}\epsilon}
$ and to write:
\begin{equation}
\lim_{\beta\to+\infty}A(\beta)=\lim_{\underset{M\to +\infty}{b\to
    0}}\left[
\prod_{m=1}^M\delta(f(\boldsymbol r_i(t_m),\boldsymbol r_j(t_m)))
\right]\label{limbetainf}
\end{equation}
Here we have exploited the fact that, if the limit for $\epsilon\to 0$
of a function $g_m(\epsilon)$ exists for every $m=1,\ldots,M$,
i.~e. $\lim_{\epsilon_M\to0}g_m(\epsilon)=g_m(0)$ with $g_m(0)\ne0$
and $|g_m(0)|<+\infty$, then we may write
the following relations:
\begin{eqnarray}
g_1(0)g_2(0)\cdots g_M(0)&=&\lim_{\epsilon_1\to 0}g_1(\epsilon_1)
\lim_{\epsilon_2\to 0}g_2(\epsilon_2)\cdots
\lim_{\epsilon_M\to 0}g_M(\epsilon_M)\nonumber\\
&=&\lim_{\epsilon\to0}(g_1(\epsilon)g_2(\epsilon)\cdots
g_M(\epsilon)) 
\end{eqnarray}
The right hand side of Eq.~(\ref{limbetainf}) is the definition of the
functional delta function $\delta(f(\boldsymbol r_i(t),\boldsymbol r_j(t)))$
which, at each instant $t$, is concentrated in the points for which
$f(\boldsymbol r_i(t),\boldsymbol r_j(t))=0$.
In other words:
\begin{equation}
\lim_{\beta\to+\infty}A(\beta)=\delta(f(\boldsymbol r_i(t),\boldsymbol r_j(t)))
\end{equation}
Remembering the definition of $A(\beta)$ of Eq.~(\ref{defabeta}), we
obtain:
\begin{equation}
\lim_{\beta\to+\infty} C(\beta) \exp\left[a\beta\int_0^{t_f}
dtf^2(\boldsymbol r_i(t),\boldsymbol r_j(t))
\right]=
\delta(f(\boldsymbol r_i(t),\boldsymbol r_j(t)))
\end{equation}
This proves Eq.~({\ref{delide}}) apart from the infinite normalization
constant $C(\infty)$.

Alternatively, one can show Eq.~(\ref{delide}) using the Fourier
representation of the Dirac delta function:
\begin{equation}
\delta(f(\boldsymbol r_i(t,s),\boldsymbol r_j(t,s)
))=\lim_{\alpha\to+\infty}\int{\cal D}
\lambda(t,s)e^{-i\int_0^{t_f}dt\int_0^Lds \lambda(t,s)f(\boldsymbol
  r_i(t,s), \boldsymbol
  r_j(t,s))}e^{-\int_0^{t_f}dt\int_0^Lds\frac{\lambda^2(t,s)}{2\alpha} }
\end{equation}
Performing the simple gaussian integral in $\lambda(t,s)$ we obtain:
\begin{equation}
\delta(f(\boldsymbol r_i(t,s),\boldsymbol r_j(t,s)))=
\lim_{\alpha\to+\infty}\int{\cal
  D}\lambda'(t,s)e^{-\int_0^{t_f}dt\int_0^Lds\frac{\lambda^{\prime
      2}(t,s)}{2\alpha} }
e^{-\frac\alpha2\int_0^{t_f}dt\int_0^Ldsf^2(\boldsymbol r_i(t,s),\boldsymbol r_j(t,s))}
\end{equation}
where $\lambda'$ is the shifted field
$\lambda'(t,s)=\lambda(t,s)+i\alpha
f(\boldsymbol r_i(t,s),\boldsymbol r_j(t,s))$.
Thus, apart from an irrelevant constaint, we may write the relation:
\begin{equation}
\delta(f(\boldsymbol r_i(t,s),\boldsymbol r_j(t,s)))=
\lim_{\alpha\to+\infty}
e^{-\frac\alpha2\int_0^{t_f}dt\int_0^Ldsf^2(\boldsymbol r_i(t,s),\boldsymbol r_j(t,s))}
\end{equation}
which coincides exactly with Eq.~(\ref{delide}).

\end{document}